\documentclass[jpcm,amsmath,amssymb]{iopart}
\newcommand{\urwchancery}[1]{{\fontfamily{pzc}\selectfont #1}}
\usepackage{float}
\usepackage{graphicx,amsfonts}
\usepackage{cite}
\usepackage{wasysym}
\usepackage[normalem]{ulem}
\usepackage{bm}
\usepackage{graphicx}
\usepackage{lmodern}
\usepackage[colorlinks, allcolors=blue]{hyperref}
\usepackage[usenames, dvipsnames]{color}
\usepackage{doi}
\begin{document}
\onecolumn
\title[]{Quadrupolar Phases and Plateau States in Skewed Ladders}
\author{Sambunath Das$^{1,2,(a),}$\footnote{Current address: Institute of Physics of the Czech Academy of Sciences, Na Slovance 1999/2, 182 00 Prague 8, Czech Republic}, Dayasindhu Dey$^{1,3,(b)}$, \\
Manoranjan Kumar$^{2,(c)}$ and S. Ramasesha$^{1,(d)}$}
\address{$^1$ Solid State and Structural Chemistry Unit, Indian Institute of Science, Bangalore 560012, India}
\address{$^2$ S. N. Bose National Centre for Basic Sciences, Block JD, Sector III, Salt Lake, Kolkata 700106, India}
\address{$^3$ UGC-DAE Consortium for Scientific Research, University Campus, Khandwa Road, Indore 452001, India.}
\ead{(a)sambunath.das46@gmail.com (b)dayasindhu.dey@gmail.com\\
     \qquad\quad(c)manoranjan.kumar@bose.res.in (d)ramasesh@iisc.ac.in}
\begin{abstract}
Two legged skewed spin-$\frac{1}{2}$ ladders are frustrated and exhibit exotic quantum phases
in ground state due to  strong quantum fluctuations and competing spin
exchanges. Here, we study ground state properties of a
spin-$\frac{1}{2}$ Heisenberg model on 3/4, 3/5 and 5/5 skewed ladders in the presence of a Zeeman magnetic field, $B$,
using exact diagonalization and the density matrix renormalization
group method. We note the existence of plateaus at $m =$ 1/3 and 2/3 for 3/4 skewed ladder,
at $m =$ 1/4, 1/2, and 3/4 for 3/5 skewed ladder, and at $m =$ 0, 1/3, and 2/3 for 5/5 skewed
ladder, where $m$ is the ratio of the observed magnetization ($M$)
to the saturated magnetization (${M_\mathrm{max}}$). The plateau state is always a gapped state
and the plateau width depends on the gap in the system. Surprisingly, the 3/4 and 5/5 skewed
ladders show interesting quadrupolar or n-type spin nematic phases below the 1/3$^{rd}$ plateau,
i.e, at very low magnetic fields. These two systems are unique as they host both a plateau and
a quadrupolar phase at low magnetic fields. The linear variation of pitch angle of the spin
with magnetization and behavior of  binding energy of magnon pairs as function of magnetic
field are also calculated in both the systems. We also study the contribution of the binding
energy to two magnon condensate.
\end{abstract}

\ioptwocol

\section{Introduction}
Frustrated low dimensional magnets have attracted a great deal of attention of the condensed matter
community due to their intriguing ground state (gs) properties, these systems may exhibit a plethora
of exotic quantum phases~\cite{ckm69a,ckm69b,hamada88,chubukov91,chitra95,white96,itoi2001,
Anderson1958,Abrahams1979} which may have potential for applications in spin based technologies.
The frustration can arise either due to geometrical arrangement of spins or
competing exchange interactions~\cite{mahdavifar2008,sirker2010,mk2015,soos-jpcm-2016,mk_bow,
chubukov91,mk2012,mk2015,vekua2007,hikihara2008,sudan2009,dmitriev2008,meisner2006,
meisner2007,meisner2009,aslam_magnon,kecke2007}. The simplest interaction driven
frustrated model is the Heisenberg spin-$\frac{1}{2}$ $J_1-J_2$ model in one-dimension (1D) where
$J_1$ and $J_2$ are the nearest and the next nearest neighbor spin exchange interactions;
antiferromagnetic $J_2$ exchange interaction induces frustration irrespective the
nature of $J_1$ interaction~\cite{chitra95,white96,itoi2001,soos-jpcm-2016,sirker2010}. The
competing nearest and next nearest exchange interaction leads to many interesting gs quantum
phases characterized by quasi-long range gapless spin liquid~\cite{chitra95,white96}, gapped
short range dimer~\cite{chitra95,white96}, spiral~\cite{chitra95,white96,itoi2001,mk2012,mk2015}
and ferromagnetic phase \cite{itoi2001,soos-jpcm-2016} etc. For ferromagnetic $J_1$
and antiferromagnetic $J_2$ this model shows a topological gs
for $|J_2/J_1| > 0.25$~\cite{agrapidis2019}.

The 1D isotropic Heissenberg $J_1-J_2$ model also known as the zigzag ladder has frustrated singlet gs for
antiferromagnetic $J_2$ regardless the
sign of $J_1$. The zigzag ladder can be conveniently represented with odd and even numbered sites
forming the two legs~\cite{chitra95}. The inter-leg interactions
are denoted by $J_1$, while the intra-leg interactions are denoted by $J_2$. For ferromagnetic
$J_1$ and in the presence of Zeeman magnetic field $B$ the
frustrated model systems exhibit
varieties of new quantum phases like the vector chiral ~\cite{hikihara2008,sudan2009,dmitriev2008,
meisner2006,aslam_magnon}, quadrupolar, hexapolar and so on in large $B$ limit \cite{hikihara2008,
sudan2009,aslam_magnon}; some of these phases like the quadrupolar phase is claimed
to have been observed experimentally ~\cite{mourigal2012}.

There are various types of antiferromagnetically coupled Heisenberg
spin-$\frac{1}{2}$ $u$/$v$ ladders, where adjacent rings
with $u$ and $v$ vertices form
a ladder-like structure. Depending on the values of $u$ and
$v$, these ladders can be
classified as 5/7, 3/4, 5/5, and 3/5 skewed ladders. These structures are called skewed
ladders due to the slanted rung bonds in the system. They
can be constructed by periodically removing some of the rung bonds of the zigzag
ladder~\cite{thomas2012,geet} as shown in Fig.~\ref{fig:schematic}.
The study of 5/7 skewed ladder was inspired by the fused azulene system
made up of 5- and 7-membered carbon rings alternately fused to yield
ladder like structure and model calculation show the ferrimagnetic gs ~\cite{thomas2012}.
The fused 5/7 membered ring structures can be realised at the grain
boundary of graphene and also in the fused azulene systems~\cite{Huang2011,
Kochat,Balasubramanian2019}. These systems, we believe, can also be realised in inorganic
supramolecular structures.

Study of short oligomers of fused azulenes using both unrestricted DFT technique
and spin models on finite fused azulene lattice revealed a triplet gs for systems of more
than eleven unit cells~\cite{Qu}. Rano et al. used ab initio techniques to look for triplet ground
states in a related system called fused acene-azulene systems~\cite{Rano}. There is also
considerable theoretical work on creating a magnetic gs in systems based on hydrocarbons
which resemble skewed ladders~\cite{Valentim_2020,Valentim_2022,Rano,Chiappe_2015}.

The 3/4 ladder can be mapped to interacting trimer system where each triangle
can be viewed as a spin trimer with next nearest neighbor interactions. In different limit of
$J_1-J_2$ parameter space, 3/4 ladder represents various coupled trimer systems which can be realised
in real materials like distorted azurite systems if the distortion results in second neighbor
interaction between end spins in the trimers~\cite{Filho_trimer,Kikuchi_trimer}
and X$_2$Cu$_3$Ge$_4$O$_{12}$ (where X is Na or K)~\cite{Bera2022,Stoll_trimer}. The gs of the Heisenberg antiferromagnetic
(HAF) spin-$\frac{1}{2}$ model on 5/7, 3/4 and 3/5 skewed ladders exhibits interesting magnetic and
non-magnetic quantum phases in the $J_{1}/J_{2}$ parameter space, whereas the 5/5 ladder remains
non magnetic across the entire parameter space. Here $J_1$ is the nearest neighbor exchange
between spins on the rung, while $J_2$ denotes the next-nearest neighbor exchange along the
leg ~\cite{geet}. The precise phase boundary between the magnetic and non-magnetic regions can
also be determined using both the entanglement entropy and fidelity calculations~\cite{Sambu_entropy}.
The Heisenberg $J_1-J_2$ spin-1 model, similar to the spin-$\frac{1}{2}$ model, on 3/4, 3/5 and 5/7 skewed
ladder geometries show interesting non-magnetic and magnetic phases, and gs exhibit vector
chiral phase on the 3/5 and 5/7 geometries~\cite{Sambu_spn1_57,Sambu_spn1_34}.

In presence of the magnetic field $B$, the gs of the $J_1-J_2$ spin-$\frac{1}{2}$ model on zigzag
and skewed ladder exhibits many interesting quantum phases. The magnetization $M$ and magnetic field
$B$ curve of this model on the zigzag ladder shows a 1/3-plateau phase for $J_2/J_1 > 0.6$ for
antiferromagnetic $J_1$ and $J_2$~\cite{okunishi_jpsj2003}. An energy gap between two consecutive
magnetic spin sectors in the thermodynamic limit results in a magnetization plateau; for example in
an integer spin HAF chain with periodic  boundary condition where the
energy gap (Haldane gap) between the gs ($S=0$) and next magnetic excited state ($S=1$) is finite in
the thermodynamic limit leads to a magnetization plateau at $m=0$,
where $m=M/M_{\mathrm{max}}$, and $M$ and $M_{\mathrm{max}}=NS$ ($N$ is
number of spins in the system and $S$ is spin at each site,) denote total
magnetization and  saturation magnetization \cite{haldane83a,haldane83b,affleck86}. The plateau
at $1/3$ magnetization are quite common in real materials for example the trimer spin-$\frac{1}{2}$
chains Cu$_3$(P$_2$O$_6$OH)$_2$~\cite{hase2006} and Na$_2$Cu$_3$Ge$_4$O$_{12}$~\cite{Bera2022} show
only one plateau phase at $m=1/3$. The magnetization plateaus at $m=1/3$ in $J_1-J_2$ type
frustrated spin-$\frac{1}{2}$ chains are realised in Cu$_3$(CO$_3$)$_2$(OH)$_2$~\cite{Kikuchi_trimer,kikuchi2006,gu2006} where the plateaus are found at $m=1/3$. Other compounds showing 1/3 plateau are
Ca$_3$Co$_2$O$_6$~\cite{zhao2010,maignan2004,hardy2004}, Sr$_3$Co$_2$O$_6$~\cite{wang2011},
Sr$_3$HoCrO$_6$~\cite{hardy2006}, SrCo$_6$O$_{11}$~\cite{ishiwata2005} and CoV$_2$O$_6$~\cite{yao2012,lenertz2011,he2009}, while the frustrated ladder compound, NH$_4$CuCl$_3$, shows two plateaus at
$m = 1/4$ and 3/4~\cite{shiramura98}. Interestingly, $J_1-J_2$ spin-$\frac{1}{2}$ model on 5/7
skewed ladders also exhibits plateaus at $m = 1/4, 1/2$ and 3/4~\cite{57plateau}.

To understand the plateau phase, Oshikawa, Yamanaka and Affleck (OYA) formulated a necessary
condition as $p (S - m) \in \mathbb{Z}$ for the occurrence of plateau in a spin-$S$ 1D system
where $S$ is the spin at each site, $p$ is the
period of the magnetic unit cell of the gs, $m$
is the magnetization of the plateau phase measured in the unit of saturation magnetization
${M_\mathrm{max}}$ and $\mathbb{Z}$ is a set of positive integers~\cite{OYA_97}.
The 1/3 plateau of $J_1-J_2$ spin-$\frac{1}{2}$ model and spin trimers obey the OYA condition.
The OYA condition is further generalised as $ n \, S \, p \, (1 - m) \in \mathbb{Z} $
for $n$ leg ladders~\cite{Cabra1997,Cabra1998}. The Haldane chain is a special case with $n = p = 1$
and  integer $S$ chains shows plateau at $m = 0$ ~\cite{haldane83a,haldane83b,affleck86}.
In a majority of cases this condition is valid \cite{meisner2007,hase2006,shiramura98}.
It also needs to be emphasized that the OYA condition is only a necessary
condition and a numerical study is essential to establish the existence of plateaus in a system.

The stabilization of the metamagnetic or multipolar phase in the gs of the $J_1-J_2$
spin-$\frac{1}{2}$ model with ferromagnetic $J_1$ in finite $B$ is an intriguing phenomenon
~\cite{hikihara2008} and it is characterized by
the presence of elementary magnetization step sizes $\Delta M > 1$ in the $M-B$
curve. Chubukov showed the quadrupolar (QP) phase is stabilized due to condensation of two
magnons to form a composite boson ~\cite{chubukov91,hikihara2008}. The order
$q=1,2, 3 ...$ of gs multipolar phases corresponds to the number of condensing magnons and the order
$q$ in the $J_1-J_2$ spin-$\frac{1}{2}$ model can be tuned by varying $J_2/J_1$ ~\cite{lauchili2009}
at high $B$. The nature of the multipole orders were analysed near the critical point
$|J_2/J_1| = 0.25$ and it was shown that close to the critical point a large number of magnons
condense with very small binding energy \cite{aslam_magnon}. Parvej and Kumar suggested that the QP
phase in the spin-$\frac{1}{2}$ $J_1-J_2$ model can be characterized by using the inelastic neutron
structure factor \cite{aslam_magnon}. In this phase the changes in magnetization in the $M-B$ curve,
$\Delta M = 2$ \cite{hikihara2008,aslam_magnon,lauchili2009}. The condensation of magnons in QP
phase ($q=2$)  is analogous to electrons forming Cooper pairs
in superconductors, except that in  the QP phase the two magnons are bosons and condense to form a
two magnon bound state. There are several reports on the detection of the QP phase,
specially in LiCuVO$_{4}$~\cite{mourigal2012,enderle2005}. The QP phase in low-dimensional systems
generally exists in the presence of ferromagnetic spin exchange interaction and in strong magnetic
field, and to the best of our knowledge it is absent in a one dimensional or ladder systems with
only antiferromagnetic spin exchange and at high magnetic field ~\cite{mourigal2012,enderle2005}.
However,finite size calculation shows that Heisenberg spin-$\frac{1}{2}$ model on two dimensional
Kagome lattice has both a plateau and steps of $\Delta M =2$ in the $M$ vs $B$
curves~\cite{schnack2020}. Several pertinent questions concerning the
3/4, 3/5, and 5/5 skewed ladders arise. For instance, how do ground-state properties of these
systems change in the presence of a magnetic field? Previous studies on the 5/7 skewed ladder revealed interesting
plateau phases in the magnetization versus magnetic field ($M-B$) curves~\cite{57plateau}. Given the distinct ground
states of different skewed ladders, it is interesting to investigate whether the 3/4, 3/5, and 5/5 systems exhibit
similar plateau features in their $M-B$ curve and whether they can give rise to different phases such as the quadrupolar phases at low magnetic
fields, considering that only antiferromagnetic exchanges are present in the spin-$\frac{1}{2}$ model.

In this paper, we  study quantum phases of HAF $J_1-J_2$ spin-$\frac{1}{2}$ model system on
3/4, 5/5 and 3/5  skewed ladders in the presence of a magnetic field and also as a function of the
ratio of rung to leg exchanges $J_1/J_2$. In all our studies we have fixed $J_2$ at unity and
it defines the energy scale. The 3/4 ladder system shows a broad plateau at 1/3 of the saturation
magnetization ${M_\mathrm{max}}$ for $J_1< 1.58$ in the presence of $B$, however, for larger $J_1$
the system shows a magnetic plateau at $m=1/3$ for $B=0$ as the gs is a high spin state and a
small 2/3 plateau  appears for $0.3<J_1<0.7$. Similarly, 5/5 ladder shows a large plateau at
$m=1/3$ for the whole range of parameters, whereas small plateaus appear at $m=$ 0 and 2/3 for $0.5< J_1 < 1.8$ and
$0.5< J_1 < 1.1$. Our third system is the 3/5 ladder which also shows three plateaus at $m$ = 1/4, 1/2 and 3/4.

The relevance of the OYA rule to plateau phases in all three systems are studied
and we show that all these plateaus follow the OYA
condition. We analyse the gs of plateau phases and show the schematic diagrams of the spin
arrangements. We also show the existence of the QP in 3/4 and 5/5 ladders below 1/3 magnetization
plateau i.e at very low magnetic Zeeman field. The QP
phase at low $B$ is another interesting and rare phenomenon in low dimensional
systems. The binding energy of the two magnon bound state is also analysed as function of the
magnetization $M$ and we show that the pitch angle follow a linear relation with $m$.

This paper is divided into five sections. In section~\ref{sec:model} we discuss the model
Hamiltonian and the numerical methods. The results for the plateau states in
3/4, 5/5, and 3/5 skewed ladders are presented in section~\ref{sec:plateau}.
In the section~\ref{sec:QP} the results for the quadrupolar phase in 3/4 and 5/5 skewed
ladders are discussed. Section~\ref{sec:disc} provides a summary  of
results and conclusions.

\section{\label{sec:model}Model and Numerical Methods}
\begin{figure}
\begin{center}
\includegraphics[width=3.0 in]{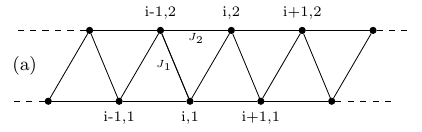}\\
\vspace{0.7cm}
\includegraphics[width=3.0 in]{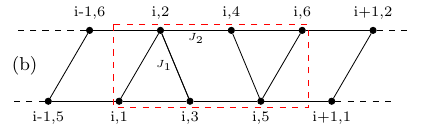}\\
\vspace{0.7cm}
\includegraphics[width=3.0 in]{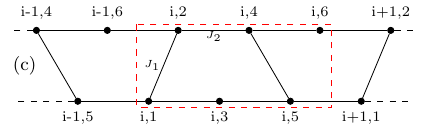}\\
\vspace{0.7cm}
\includegraphics[width=3.0 in]{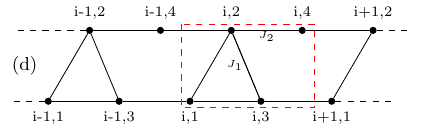}
\caption{\label{fig:schematic}Schematic diagram of (a) the regular
zigzag chain, (b) 3/4 skewed ladder: The
nearest neighbor or rung interaction is $J_1$ and the next nearest neighbor
(along the leg) interaction is $J_2$. (c) 5/5 skewed ladder and (d) 3/5 skewed ladder.
Here `i' is the index of the unit cell and the numerals 1, 2, \ldots are numbering
of the spins within the unit cell. There are 6 spins per unit cell in the 3/4 and 5/5 ladder
whereas there are 4 spins per unit cell in 3/5 ladder. The sites on the
top leg are even numbered and on the bottom leg are odd numbered.}
\end{center}
\end{figure}

The skewed ladders and the associated site numbers are shown Fig.~\ref{fig:schematic} for the 
3/4, 5/5 and 3/5 systems. The sites are numbered such that odd numbered sites are on the bottom 
leg and even numbered sites are on the top leg. Thus the rung bonds
are the nearest neighbor exchanges $J_1$ and the bonds on the legs are the next nearest
neighbor exchanges $J_2$=1. The spin value at each site is $\frac{1}{2}$. The model Hamiltonian of 
the 3/4 skewed ladder in a magnetic field is written as 
\begin{eqnarray}
	H_{3/4} &=& J_1 \sum_{i=0}^n \left[\left(\vec{S}_{i,1} + \vec{S}_{i,3}\right)\cdot \vec{S}_{i,2} +
	\left(\vec{S}_{i,4} + \vec{S}_{i,6} \right) \cdot \vec{S}_{i,5} \right]  \nonumber \\
	& &+ J_2 \sum_{i=0}^n \bigg( \vec{S}_{i,5} \cdot
	\vec{S}_{i+1,1}
	+ \vec{S}_{i,6} \cdot \vec{S}_{i+1,2}  \nonumber \\
	& &  \qquad + \sum_{k=1}^4
	\vec{S}_{i,k} \cdot \vec{S}_{i,k+2} \bigg) -B\sum_{i=0}^n \sum_{k=1}^{6} S^z_{i,k},
\label{eq:ham34}
\end{eqnarray}
where $i$ labels the unit cell, $k$ the spins within the unit cell and $n$ is the number of unit cells
(Fig.~\ref{fig:schematic}). The first term denotes the rung exchange terms, the
second term denotes the exchange interactions along the legs and the third term 
represents the interaction of the spins within a Zeeman magnetic 
field B in units of $J_2 / g \mu_B$.
Similarly, the model Hamiltonian for the 5/5 and 3/5 systems in a magnetic field is written as
\begin{eqnarray}
	H_{5/5} &=& J_1 \displaystyle \sum_{i=0}^n \left(\vec{S}_{i,1} \cdot \vec{S}_{i,2} +
    \vec{S}_{i,4} \cdot \vec{S}_{i,5} \right) \nonumber \\
	& &+ J_2 \sum_{i=0}^{n} \bigg( \vec{S}_{i,5} \cdot
	\vec{S}_{i+1,1}
	+ \vec{S}_{i,6}\cdot\vec{S}_{i+1,2} \nonumber \\ 
	& &  \qquad + \sum_{k=1}^3
	\vec{S}_{i,k} \cdot \vec{S}_{i,k+2} \bigg) -B\sum_{i=0}^n \sum_{k=1}^{6} S^z_{i,k},
\label{eq:ham55}
\end{eqnarray}
and
\begin{eqnarray}
	H_{3/5} &=& J_1 \sum_{i=0}^n \left(\vec{S}_{i,1} \cdot \vec{S}_{i,2} + \vec{S}_{i,2}\cdot \vec{S}_{i,3}\right) \nonumber \\
	& & + J_2 \sum_{i=0}^n \bigg( \vec{S}_{i,3} \cdot \vec{S}_{i+1,1} 
        + \vec{S}_{i,4} \cdot \vec{S}_{i+1,2} \nonumber \\
	& & \qquad + \sum_{k=1}^2
	\vec{S}_{i,k} \cdot \vec{S}_{i,k+2} \bigg) -B\sum_{i=0}^n \sum_{k=1}^{4} S^z_{i,k}.
\label{eq:ham35}
\end{eqnarray}
We use exact diagonalization (ED) technique for finite ladders with up to 24
spins and exploit the symmetries by using periodic boundary condition (PBC).
In all three systems reflection symmetry is present. An extra rung is needed
when open boundary condition (OBC) is used in all three cases. For larger
system sizes we use the density matrix renormalization group (DMRG)
method~\cite{white-prl92,white-prb93,schollwock2005,karen2006}
to handle the large degrees of freedom in the many body Hamiltonian. This method is a state of the 
art numerical method and is based on systematic truncation of irrelevant degrees of freedom. We 
retain up to 600 block states $({\chi}=600)$ 
which are the eigenvectors of the block density matrix with dominant eigenvalues.
The chosen value of ${\chi}$ keeps the truncation 
error to less than $\sim 10^{-10}$. We also carry out 6-12 finite sweeps to improve convergence. 
The largest system size studied is up to 500 sites for the 3/4 ladder, 392 sites for the 5/5 
ladder and 502 sites for the 3/5 ladder systems with OBC. The spin correlations in all the three 
systems is short ranged and hence the chosen sizes are adequate to study the magnetic properties. 
The model Hamiltonian preserves the total $M_s$, therefore, the DMRG calculations are carried out 
in different $M_s$ sectors of the ladder Hamiltonian.

\section{\label{sec:plateau} Plateau states in 3/4, 5/5 and 3/5 ladders}
In this section we discuss the plateau states in three different systems
3/4, 5/5 and 3/5 in subsections \ref{sbsec:MP34}, \ref{sbsec:MP55} and \ref{sbsec:MP35}, respectively.
In each subsection we discuss the $m-B$ plots, spin arrangements in large $J_1$ coupling
limit and $B_M-J_1$ curve which  characterises the magnetic field behaviour for
different $J_1$. $B_M$ is defined as the magnetic field required to close the energy gap between 
$M_s = M$ and $M_s = M+1$ states.
For a model Hamiltonian where $M_s$ is conserved,
the lowest energy state in any $M_s$ sector can be written as a function of Zeeman 
magnetic field $B$
\begin{equation}
E(M_s,B) = E_0 (M_s, B=0) - BM_s,
\label{eq:mag_ene}
\end{equation}
where $E(M_s,B)$ and $E_0 (M_s, B=0)$ are lowest energy states in the $M_s$ sector with and without an external magnetic field $B$. $B_M$ can also be defined as $E(M_s,B_M) = E(M_{s+1}, B_M)$, and in units 
of $J_2/g\mu_B$ it is given by  
\begin{eqnarray}
        B_M&=&\frac{E(M+1)-E(M)}{g\mu_B}.
\label{eq:BM}
\end{eqnarray}
$E(M+1)$ and $E(M)$ are  lowest energies in $(M+1)^{th}$ and $(M)^{th}$ total $M_s$ sectors.
The dependence of plateau width $w_{n}=B^U_{n}-B^L_{n}$, 
(where $B^L_n$ and $B^U_n$ are the lower and 
upper critical values of the magnetic field for the $n^{th}$ plateau) 
on $J_1$, is an important parameter for any practical uses of a material.
The spin bond orders $b_{kl}$ can be defined as
\begin{eqnarray}
	b_{kl}=\frac{1}{4}-\langle \hat{S}_k.\hat{S}_l \rangle,
\label{eq:BM1}
\end{eqnarray}
it is done such that a perfect singlet has $b_{kl}=1$, where as for a perfect
triplet the bond order should be $b_{kl}=0$. We also study the spin density $\rho_{k_{i}} = \langle S^z_{k_{i}} \rangle$, 
(where $`k$' is the site index and $`i$' is the unit cell index) and bond orders $b_{kl}$ in the plateau phases. In 
order to calculate the spin density for different ladder systems, we consider only one unit cell; therefore, index $`i$' 
is omitted in what follows.

\subsection{\label{sbsec:MP34} Plateau phases in the 3/4 ladder}
\begin{figure}
\begin{center}
\includegraphics[width=3.2 in]{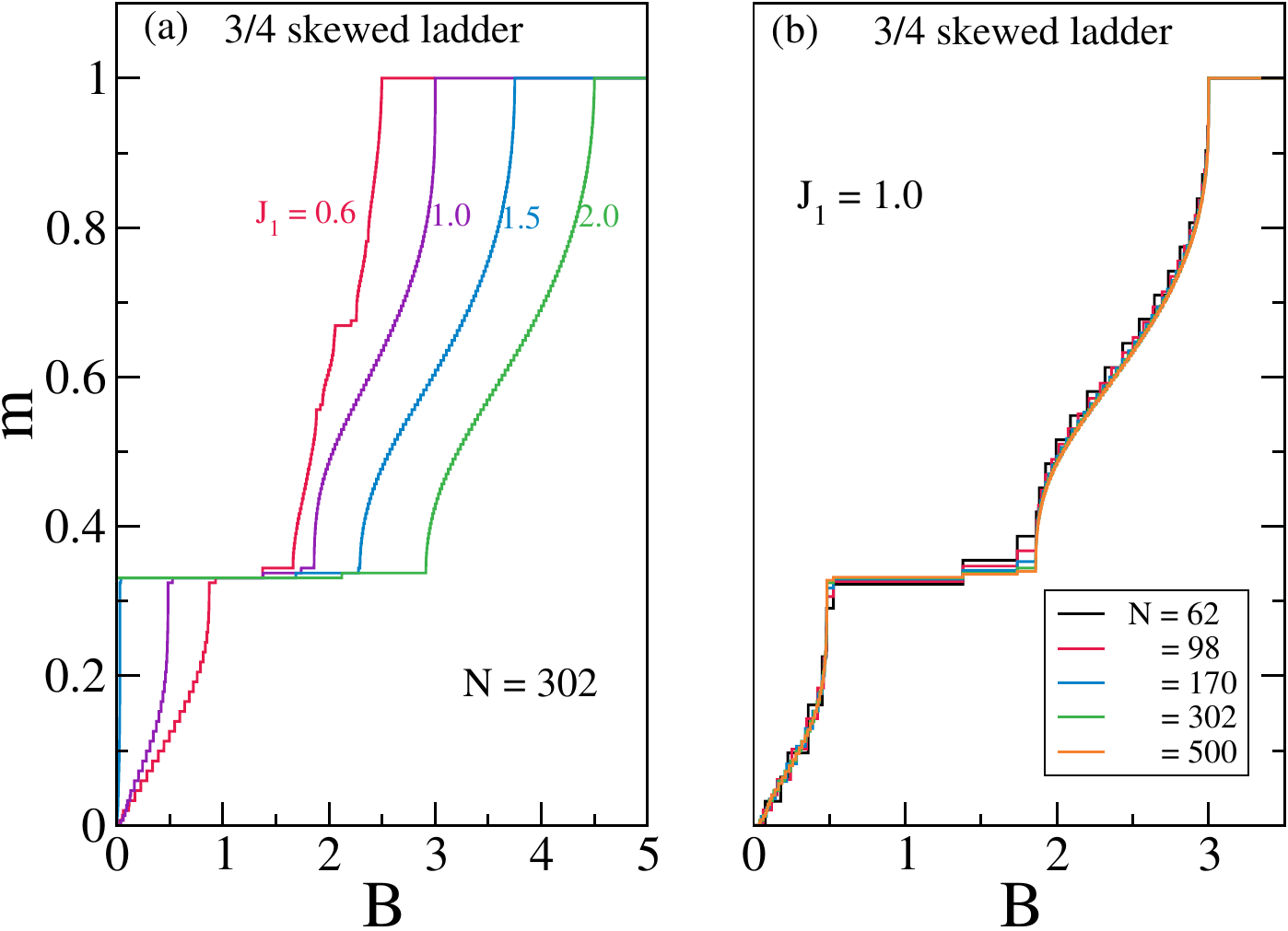}
	\caption{\label{fig:plt34} (a) $m-B$ curve for a 3/4 skewed ladder for $J_1 = 0.6$, 1.0, 1.5 and 2.0 for
$N = 302$ sites. (b) The finite size effect of the $m-B$ curve with $J_1 = 1.0$ for five
	system sizes $N = 62$, 98, 170, 302 and 500. Scale on the vertical axis is the same in both (a) and (b).}
\end{center}
\end{figure}

In this system there are 6 spins per unit cell, therefore, the OYA condition suggests 4 possible 
plateaus at $m$ = 0, 1/3, 2/3 and 1. We plot the $m-B$ curve for the 3/4 ladder for four values of 
$J_1$, namely 0.6, 1, 1.5 and 2 for $N=302$ as shown in Fig.~\ref{fig:plt34}(a). This system 
exhibits 1/3 plateau for all $J_1$ values and the gs is in 1/3 magnetic
state in the absence of $B$ for $J_1 \ge 1.58$. This system also has a small
2/3 plateau for $0.3 < J_1 <0.7$, as seen for $J_1 = 0.6$ in Fig.~\ref{fig:plt34}(a). 
The finite size effect on the plateau width is shown in Fig.~\ref{fig:plt34}(b)
for five different system sizes $N=62, \, 98, \, 170, \, 302$ and 500. 
We notice the appearance of elementary magnetization steps of $\Delta M = 2$ below $m=1/3$ plateau whereas
they change by steps of $\Delta M = 1$ above the plateau.
There is small plateau on the onset and at the end of the plateau and these 
are sensitive to finite size effects as shown in Fig.~\ref{fig:plt34}(b). The small and size 
dependent plateaus near the edge of 1/3 plateaus appear because of OBC.
\begin{figure}
\begin{center}
\includegraphics[width=3.2 in]{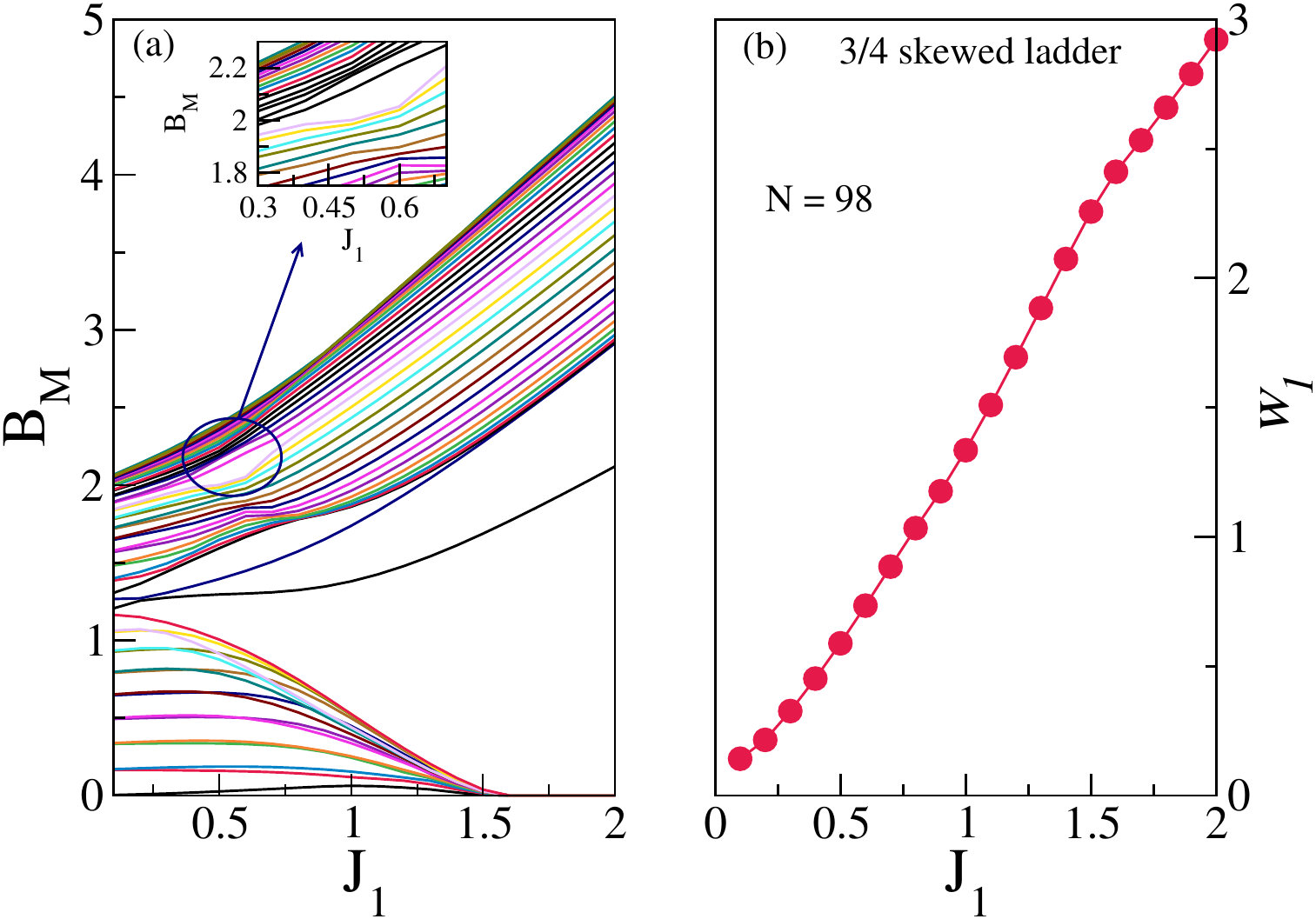}
	\caption{\label{fig:BM34} (a) The magnetic field $(B_M)$ required to close the energy 
	gap between successive lowest energy $M_s$ states vs the rung bond interaction $J_1$. 
	The inset highlights the region $0.3 < J_1 < 0.7$, providing a closer view of an 
	additional band formation appears for $m = 2/3$. (b) The width of the $m =1/3$ plateau vs. $J_1$.}
\end{center}
\end{figure}

\begin{figure}
\begin{center}
    \includegraphics[width=3.2 in]{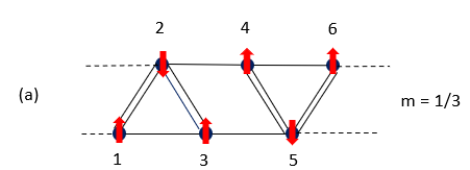}
    \includegraphics[width=3.2 in]{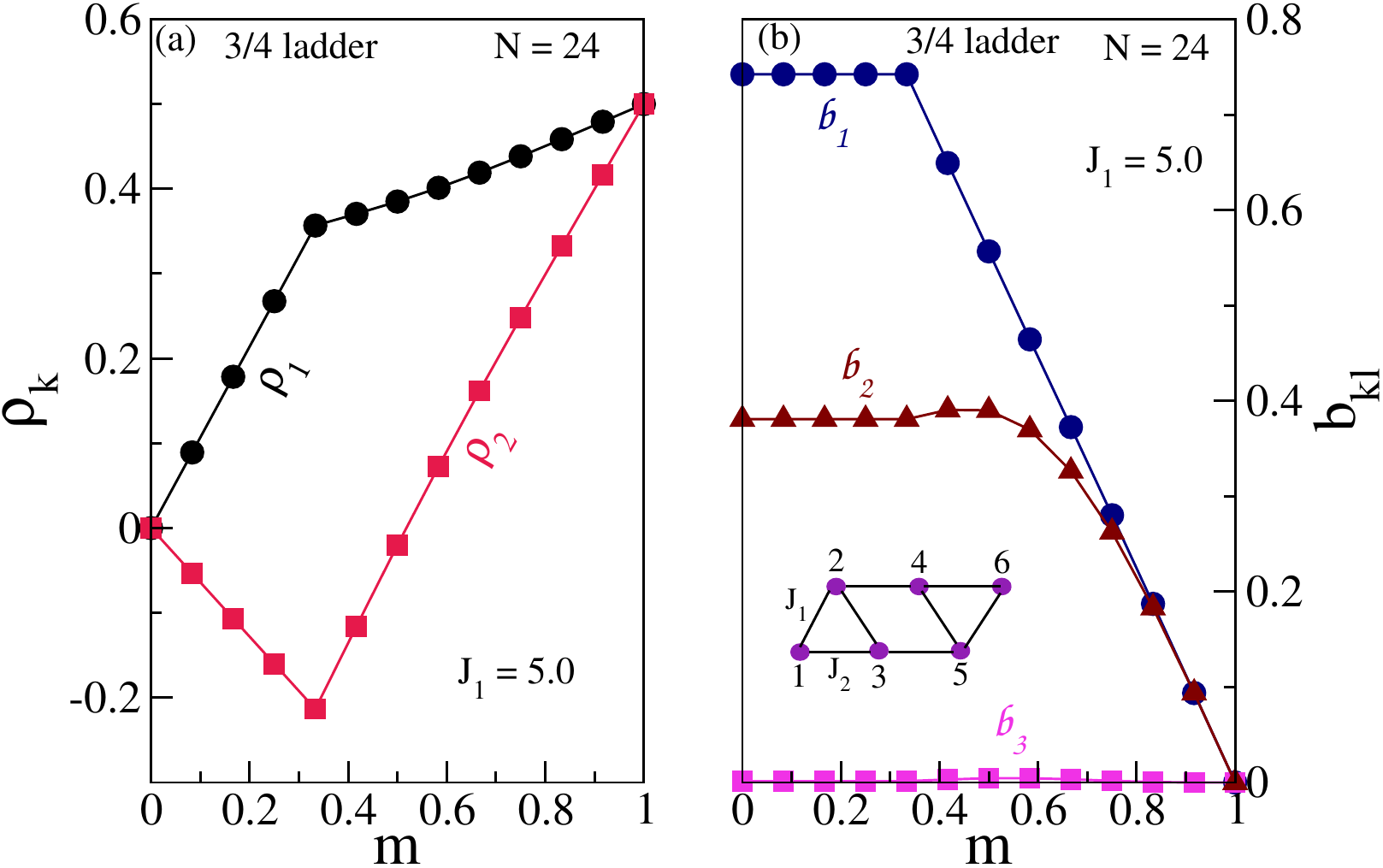}
	\caption{\label{fig:sdbo34}(a) Spin arrangements, (b) spin densities and (c) bond orders 
	in a unit cell of a 3/4 skewed ladder of $N = 24$ sites with a periodic boundary condition. 
	Here $\rho_1$ is the spin density at base sites of the triangle 
	(1, 3, 4 and 6) and $\rho_2$ is at apex sites (2 and 5). \urwchancery{b}$_1$ corresponds to the base 
	bonds ($b_{13}$ and $b_{46}$), \urwchancery{b}$_2$ is rung bonds ($b_{12}$, $b_{23}$, $b_{54}$ and 
	$b_{56}$) and \urwchancery{b}$_3$  is the bond between the apex and the base 
	sites on the same leg ($b_{24}$ and $b_{35}$).}
\end{center}
\end{figure}

In Fig.~\ref{fig:BM34} $B_M$ is plotted as a function of $J_1$ and we notice  that for small values 
of $J_1$, $B_M$ is almost equally spaced but at higher values of $J_1$, $B_M$ vs $J_1$ curves 
almost form a band. For $J_1>0$, first band corresponds to  $m = 1/3$, and second band
corresponds to saturation magnetization, $m = 1$. For $ 0.3 < J_1 < 0.7 $ additional band
formation appears for $m = 2/3$ as shown in Fig.~\ref{fig:BM34}(a). Fig.~\ref{fig:BM34}(b)
represents the width of $m =1/3$ plateau ($w_1$) as a function of $J_1$, and shows almost linear 
variation with $J_1$ in two regimes with slopes 2.059 for $J_1 < 1.5$ and 1.336 for $J_1 > 1.5$. 
We notice that the finite size effect in 1/3 plateau, $w_1$ is vanishingly small, whereas in the 
2/3 plateau it shows moderate finite size effect. The arrangement of spins in large $J_1$ limit 
is shown Fig.~\ref{fig:sdbo34}(a), and we notice that the two base spins of the triangle have
ferromagnetic alignment. The rung bonds are dominant whereas the bonds connecting
the two triangles are weak.

The spin densities and bond orders are calculated as a function of magnetization
to understand the spin configuration of the gs as well as the plateau phases.
In this system there are two types of unique spin densities: first type $\rho_1$
is spin density at base sites (1, 3, 4 and 6) and second type $\rho_2$ is spin density at apex sites
(2 and 5)(Fig.~\ref{fig:schematic}(b)). There are also three types of bond orders: first type \urwchancery{b}$_1$ is the base bonds ($b_{13}$ and $b_{46}$), second type \urwchancery{b}$_2$ is rung bonds ($b_{12}$, $b_{23}$, $b_{54}$ and $b_{56}$) and third type \urwchancery{b}$_3$ is bond between the apex and the base sites on the same leg ($b_{24}$ and $b_{35}$).
These quantities are plotted as a function of $m$ for $J_1=5$ (Fig.~\ref{fig:sdbo34}). We
notice that base sites have spin densities $\rho_1 = 0.356$  and  apex sites have  $\rho_2 = -0.215$
at 1/3 plateau which is the gs for $J_1 > 1.58$ for $B=0$. The spin densities vary linearly with 
two different slopes below 1/3 plateau and above 1/3 plateau. For $J_1 = 5$, 
\urwchancery{b}$_1$ and \urwchancery{b}$_2$ are  0.742 and 0.381 respectively for 
$m \le 1/3$ and decreases with increase in $m$. \urwchancery{b}$_3$ is vanishingly 
small i.e., the base sites of a triangle are very nearly in a triplet state on the lower leg.
\subsection{\label{sbsec:MP55} Plateau phases in the 5/5 ladder}
\begin{figure}[b]
\begin{center}
    \includegraphics[width=3.2 in]{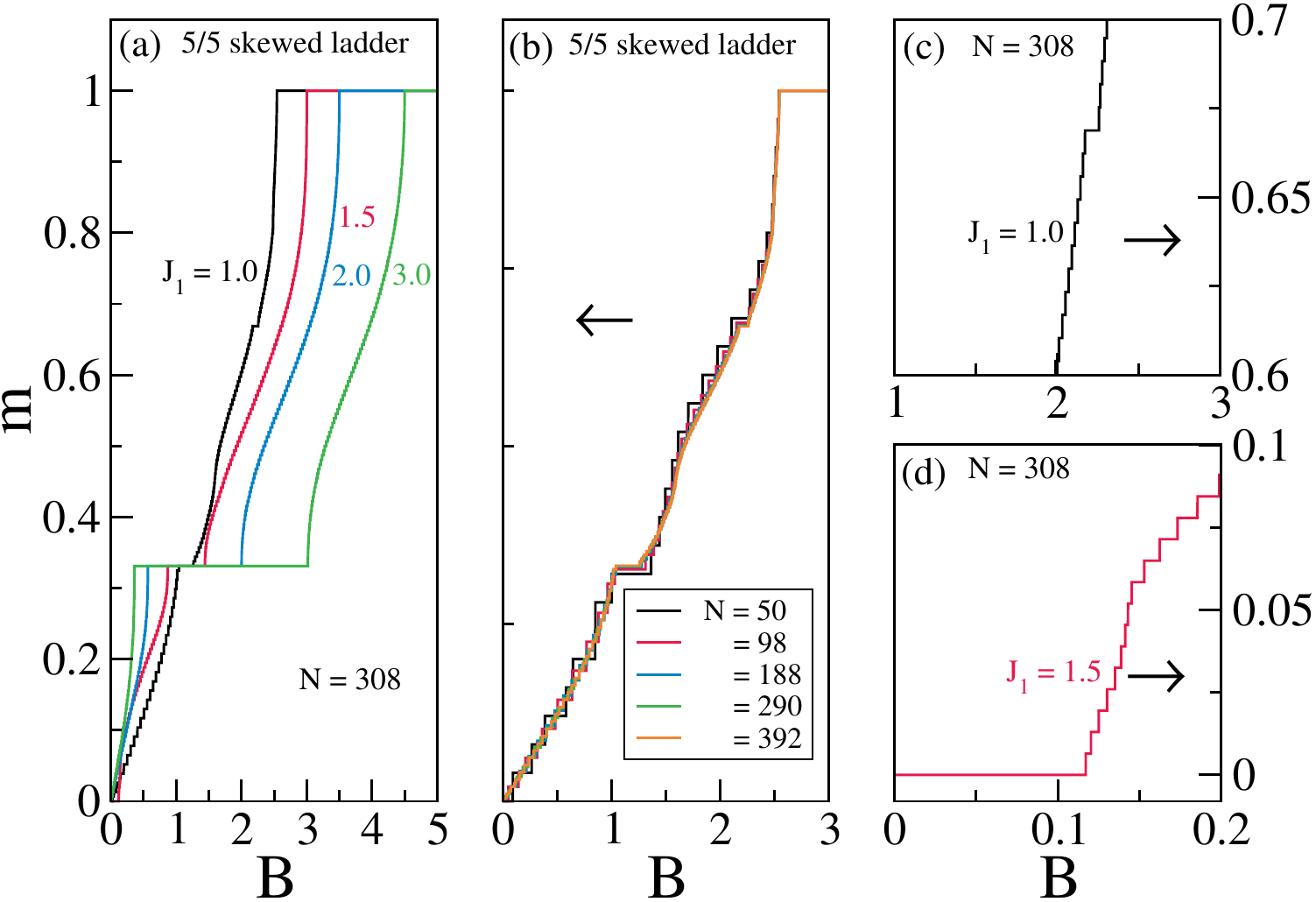}
    \caption{\label{fig:plt55}(a) $m-B$ curve for a 5/5 skewed ladder for $J_1 = 1.0$, 1.5, 2.0 
	and 3.0 for $N= 308$ sites. (b) The finite size effect of the $m-B$ curve with $J_1 = 1.0$ 
	for five system sizes $N = 50$, 98, 188, 290, 392. (c) Near the plateau at $m = 2/3$ 
	for $J_1 = 1.0$ and (d) above the plateau at $m = 0$ for $J_1 = 1.5$. Note that the 
	scale of $m$ in (c) and (d) are different from those in (a) and (b). Direction of arrows 
	indicates the scale of $m$ on the vertical axis.}
\end{center}
\end{figure}
The 5/5 ladder system also have six sites per unit cell and therefore, there are four possible
plateaus at $m$=0, 1/3, 2/3 and 1, according to the OYA criterion. In Fig.~\ref{fig:plt55}
$m-B$ curves are plotted for four values of $J_1$, namely, 1.0, 1.5, 2.0 and 3.0 for $N = 308$.
This system also exhibits a dominant $m = 1/3$ plateau in the presence of external magnetic field
$B$, besides a small 0 and 2/3 plateau for $J_1=1.5$ and  $J_1=1$,
respectively as shown in Fig.~\ref{fig:plt55}(a), (c) and (d). The finite size effect on the
plateau width is shown in Fig.~\ref{fig:plt55}(b)  for five different system sizes
$N=50$, 98, 188, 290 and 392 for $J_1=1$.
We observe the emergence of elementary magnetization steps of $\Delta M = 2$
below $m = 1/3$ plateau for $J_1=1$, while they
vary by steps of $\Delta M = 1$ above the 1/3 plateau.
Similar to the 3/4 ladder this system also shows small plateaus at the onset and at the
end of 1/3 plateau which are sensitive to system size
(Fig.~\ref{fig:plt55}(b)).
\begin{figure}[b]
\begin{center}
    \includegraphics[width=3.2 in]{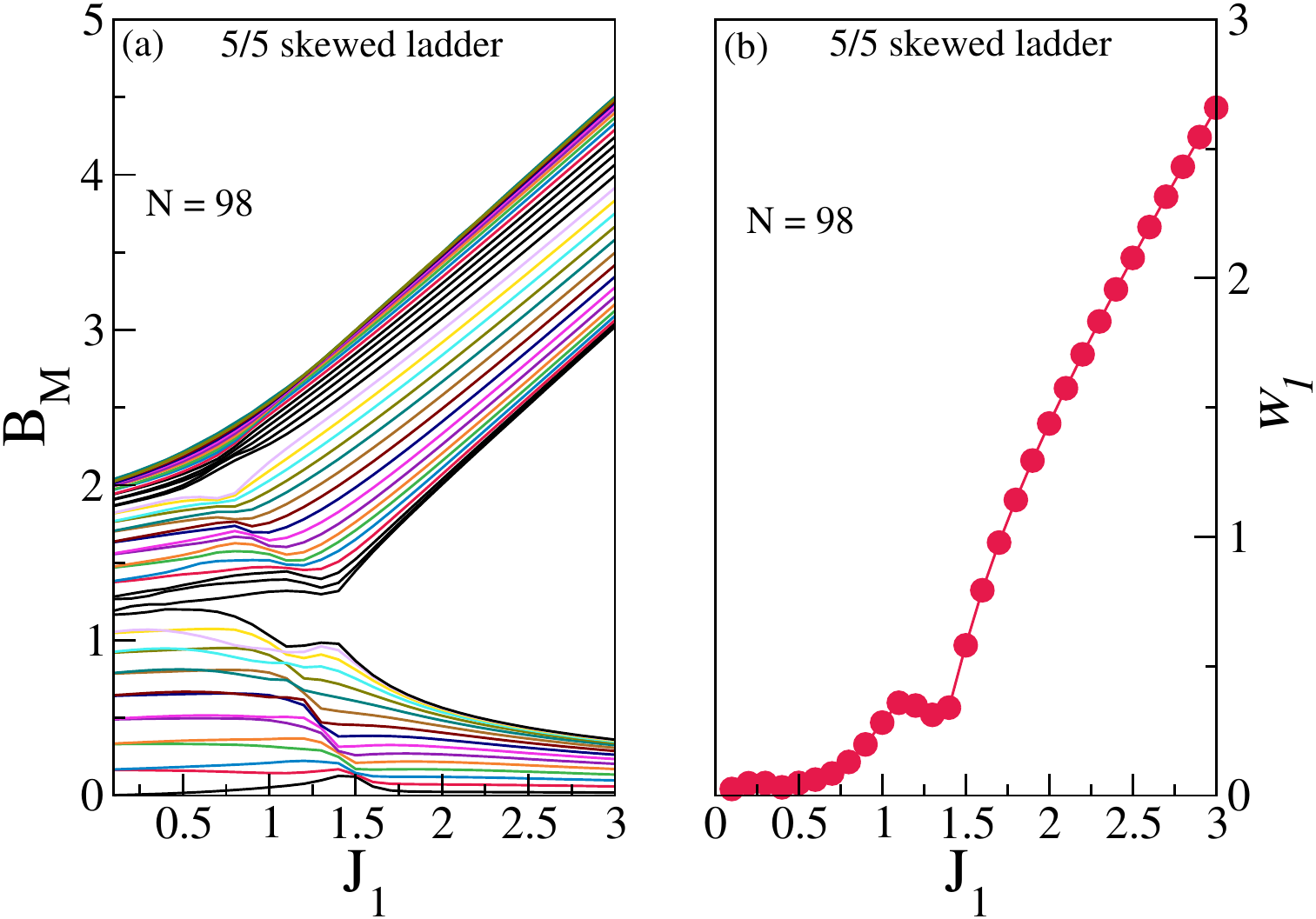}
    \caption{\label{fig:BM55}(a) The magnetic field $(B_M)$ required to close the energy gap between
successive lowest energy $M_\mathrm{s}$ states vs the rung bond interaction $(J_1)$ for a 5/5
skewed ladder. (b) The width of the  $m = 1/3$ plateau vs. $J_1$.}
\end{center}
\end{figure}

\begin{figure}
\begin{center}
\includegraphics[width=3.2 in]{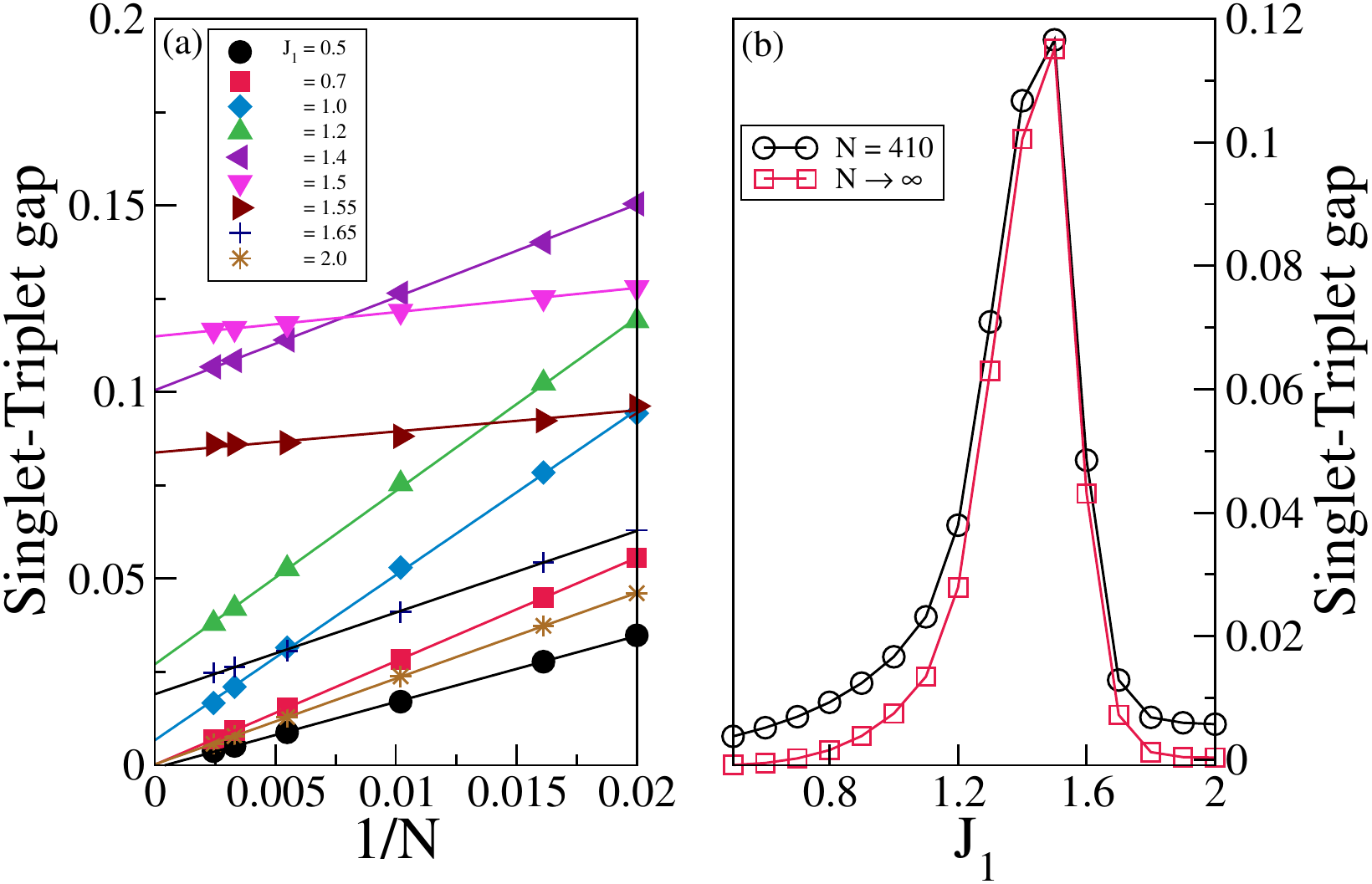}
	\caption{\label{fig:stgap_55}(a) Variation of the singlet triplet gap with
	the inverse system size $1/N$ of a 5/5 skewed ladder for different $J_1$ values, 
	(b) singlet-triplet gap for a 5/5 skewed ladder in thermodynamic limit for different 
	$J_1$ values.}
\end{center}
\end{figure}

In Fig.~\ref{fig:BM55}(a), $B_M$ is plotted as a function of $J_1$ and we notice  that
for small values of $J_1$, the $B_M$ curves are almost equally spaced and form bands at
higher values of $J_1$. For $J_1 > 0$, the first band corresponds to $m=1/3$, and
the second band corresponds to saturation magnetization for $J_1>2$.
For $ 0.5 < J_1 < 1.1 $, a gap opens in $B_M$ near the 2/3 plateau, and for the same
parameter regime, the lowest $B_M$  have finite value. A large value for lowest
$B_M$ indicates a large singlet and triplet gap. Fig.~\ref{fig:BM55}(b)
represents the 1/3 plateau width $(w_1)$ as a function of $J_1$ for $N=98$. We
notice that the width is tiny for $J_1 <0.5$ and increases slowly up to 1.
$w_1$ is almost constant for $1<J_1<1.5$ and increases linearly beyond $J_1=1.5$.
The finite size effect on $w_1$ is almost negligible for large $J_1$.
For small value of $J_1(<0.5)$ the width for both $m = 0$ and 2/3 plateaus are vanishingly 
small in thermodynamic limit. The zero magnetization plateau is due to finite 
singlet-triplet (ST) gap and is plotted
as a function of $1/N$ for various values of $J_1$ in Fig. \ref{fig:stgap_55}(a)
and the extrapolated value of ST gap is shown in Fig~\ref{fig:stgap_55}(b). The ST
gap is finite for $0.6 < J_1 < 1.8$ and has maximum at $1.5$. To understand the maxima in 
the singlet-triplet gap we focus on the spins at sites 3, 4, 5 and 6 in each unit cell. In the 
small $J_1$ limit, spins at 4 and 6 form a singlet and so will spins at 3 and 5 leaving the spins 
at 4 and 5 largely uncorrelated leading to a small spin gap. In the large $J_1$ limit,
the spins at 3 and 6 form a strong singlet, again leaving the spins 4 and 5 uncorrelated, 
resulting in a vanishing spin gap in the large $J_1$ limit. For intermediate $J_1$ values the 
crossover between these two pictures results in a maxima in the spin gap which in 
our case peaks for $J_{1} \sim 1.5$. 

\begin{figure}
\begin{center}
\includegraphics[width=3.2 in]{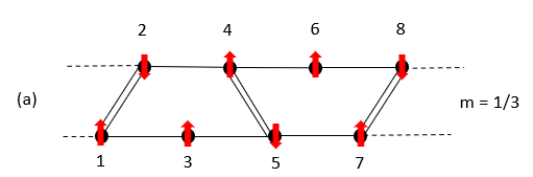}
\includegraphics[width=3.2 in]{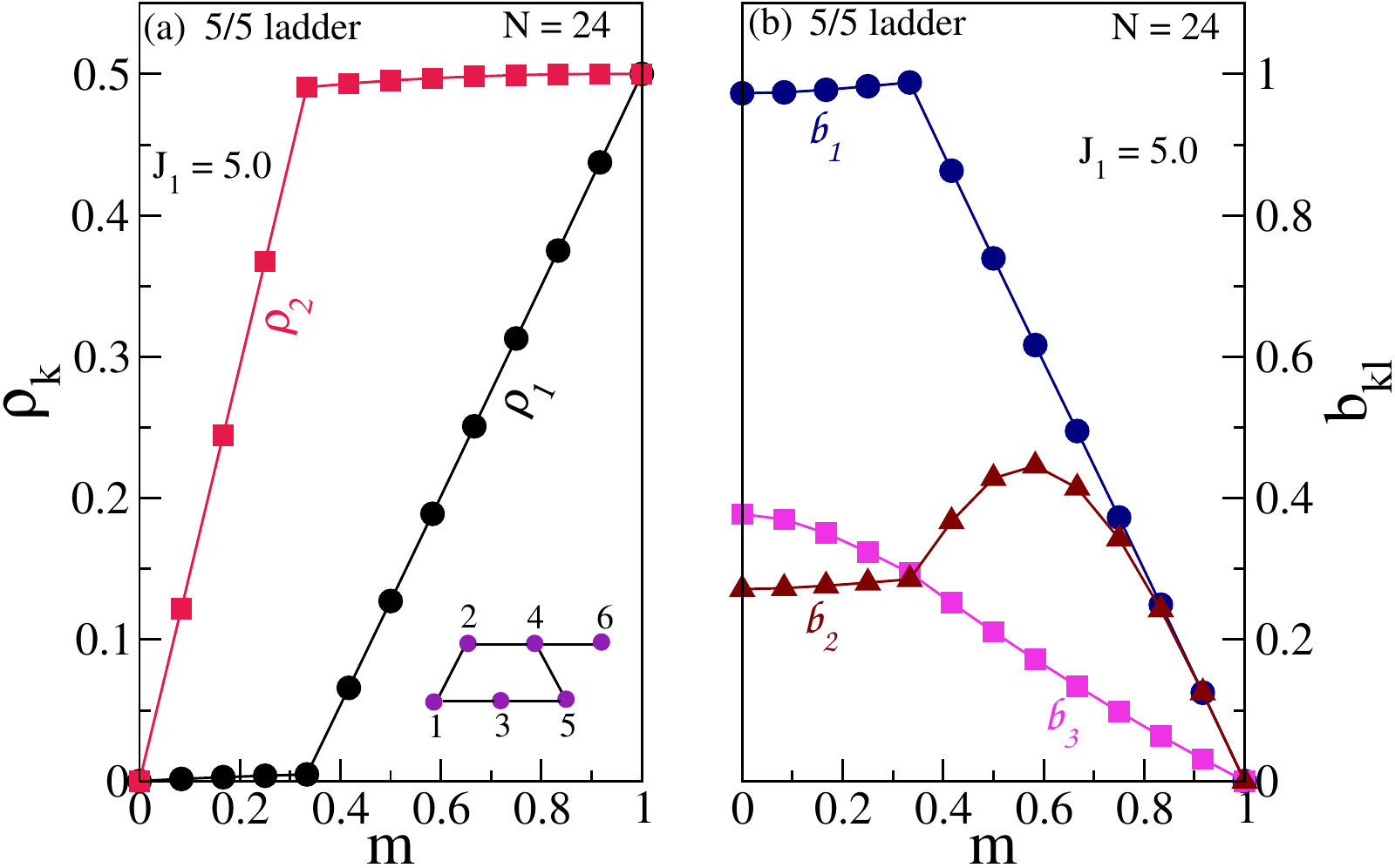}
	\caption{\label{fig:sdbo55}(a) Spin arrangements, (b) spin densities and (c) bond orders 
	in a unit cell of a 5/5 ladder of $N =24$ sites with periodic boundary condition. 
	Here $\rho_1$ is the spin density at sites (1, 2, 4 and 5) and 
	$\rho_2$ is at sites (3 and 6). \urwchancery{b}$_1$ corresponds to the rung bonds ($b_{12}$ and 
	$b_{45}$), \urwchancery{b}$_2$ connects the the singlet rung bond site and free spin site 
	($b_{13}$, $b_{35}$ and $b_{46}$) and \urwchancery{b}$_3$ corresponds to the bond between the sites 2 
	and 4 ($b_{24}$).}
\end{center}
\end{figure}

In 5/5 ladder there are six sites per unit cell and is a highly symmetric structure,
therefore, there are only two types of unique spin densities: first type $\rho_1$ is at sites 
(1, 2, 4 and 5) and second type $\rho_2$ is at sites (3 and 6) as shown in 
Fig.~\ref{fig:schematic}(c). There are three types of bond orders: first type are the rung 
bonds ($b_{12}$ and $b_{45}$) which we designate as \urwchancery{b}$_1$, second type \urwchancery{b}$_2$ 
connects the singlet rung bond site and free spin site, examples of which are $b_{13}$, $b_{35}$
and  $b_{46}$. Third type \urwchancery{b}$_3$ are bonds connecting sites of 
two nearest singlet rung bonds such as $b_{24}$ and  $b_{57}$. The spin densities and bond orders
are calculated as a function of magnetization to understand the spin configuration
of the gs in the plateau phases (Fig.~\ref{fig:sdbo55}). We note that first type of
spin has $\rho_1= 0.004$  and  second type has $\rho_2$ is 0.490 at 1/3 plateau
for $J_1=5$. Spin density $\rho_1$ increases linearly with $m$ for $ m > 0.33$. For $J_1=5$, 
the bond orders \urwchancery{b}$_2$ and \urwchancery{b}$_3$ 
are  0.285 and 0.295 and are, therefore, weakly antiferromagnetic in nature. In a unit cell all the rung bonds 
$b_{12}$ and $b_{45}$ form strong singlets, whereas, spin on sites 3 and 6 behaves like free spin. For $m>1/3$, 
\urwchancery{b}$_2$ increases and has maximum at $m \approx 0.6$ and decrease thereafter, 
whereas \urwchancery{b}$_1$ decreases as all the spins in the bond align ferromagnetically 
at saturation field.
\subsection{\label{sbsec:MP35} Plateau phases in the 3/5 ladder}
\begin{figure}[h]
\begin{center}
	\includegraphics[width=3.2 in]{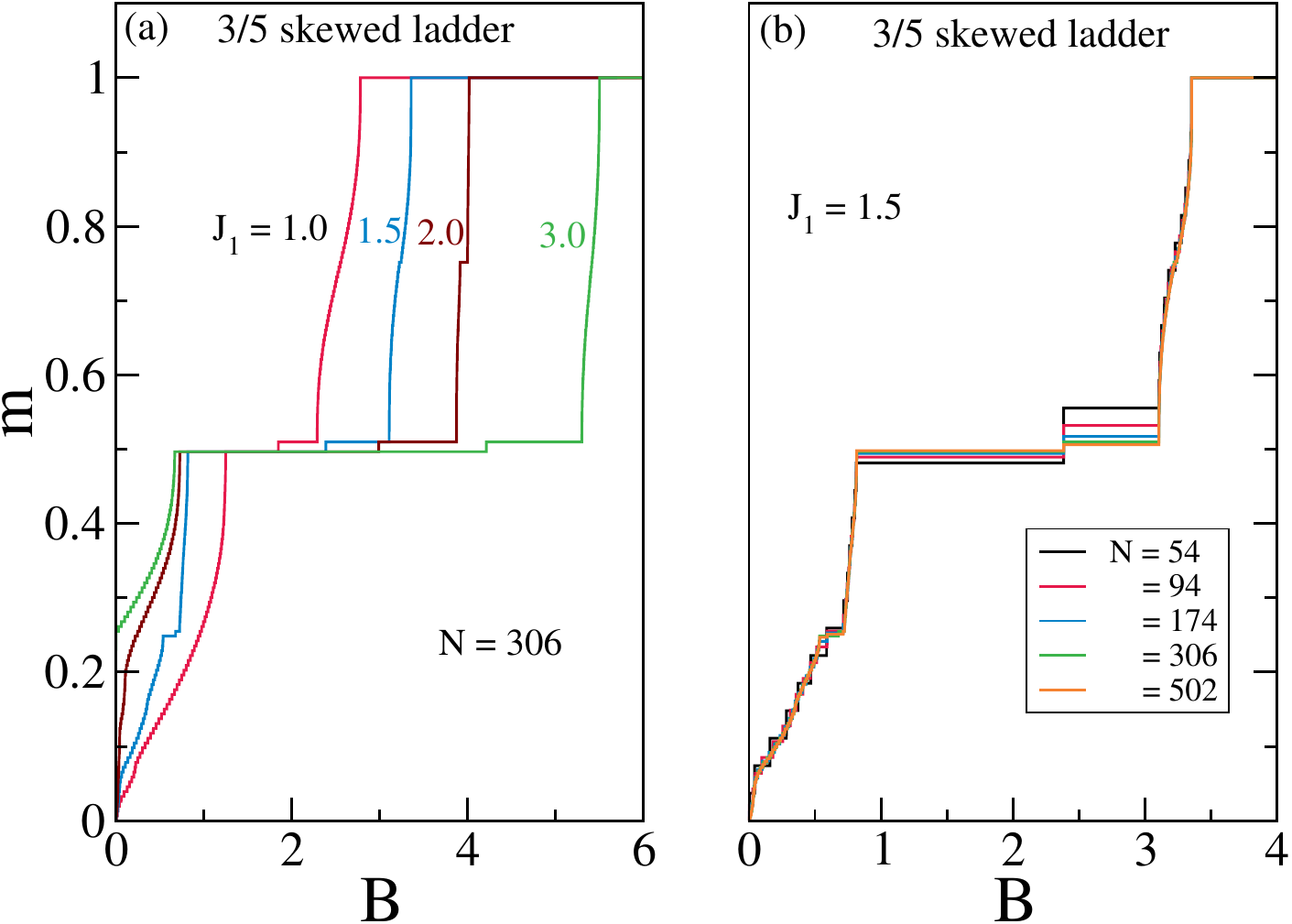}
	\caption{\label{fig:plt35}(a) $m-B$ curve for a 3/5 skewed ladder for
	$J_1 = 1.0$, 1.5, 2.0 and 3.0 for $N = 306$ sites. (b) The finite size
	effect of the $m-B$ curve with $J_1 = 1.5$ for five system sizes $N = 54$,
	94, 174, 306, and 502. Vertical scale is same as in (a).}
	\end{center}
\end{figure}

Our third system, the 3/5 skewed ladder, has four sites per unit cell as shown in 
Fig.~\ref{fig:schematic}(d). According to OYA criterion this system can have only three possible 
plateaus at $m=0$, 1/2 and 1. In Fig.~\ref{fig:plt35}(a), we show the $m-B$ curves for this system 
for four values of $J_1= 1$, 1.5, 2 and 3 for a ladder with 306 sites. This system exhibits a
dominant $m = 1/2$ plateau; besides this dominant plateau, the system also has two narrow plateaus 
at $m = 1/4$ and 3/4 for $J_1=1.5$ and $J_1 = 2.0$, respectively. Interestingly,
$m = 1/4$ becomes the gs for $J_1 > 2.3$ in zero field. The finite size effect on the plateau
width is shown in Fig.~\ref{fig:plt35}(b) for five different system sizes of $N=54$, 94,
174, 306 and 502 for $J_1 = 1.5$. Similar to the 3/4 and 5/5 ladders this system also
shows small plateaus at the onset and at the end of the 
1/2 plateau which are sensitive to the finite size of the system. 

\begin{figure}
\begin{center}
	\includegraphics[width=3.2 in]{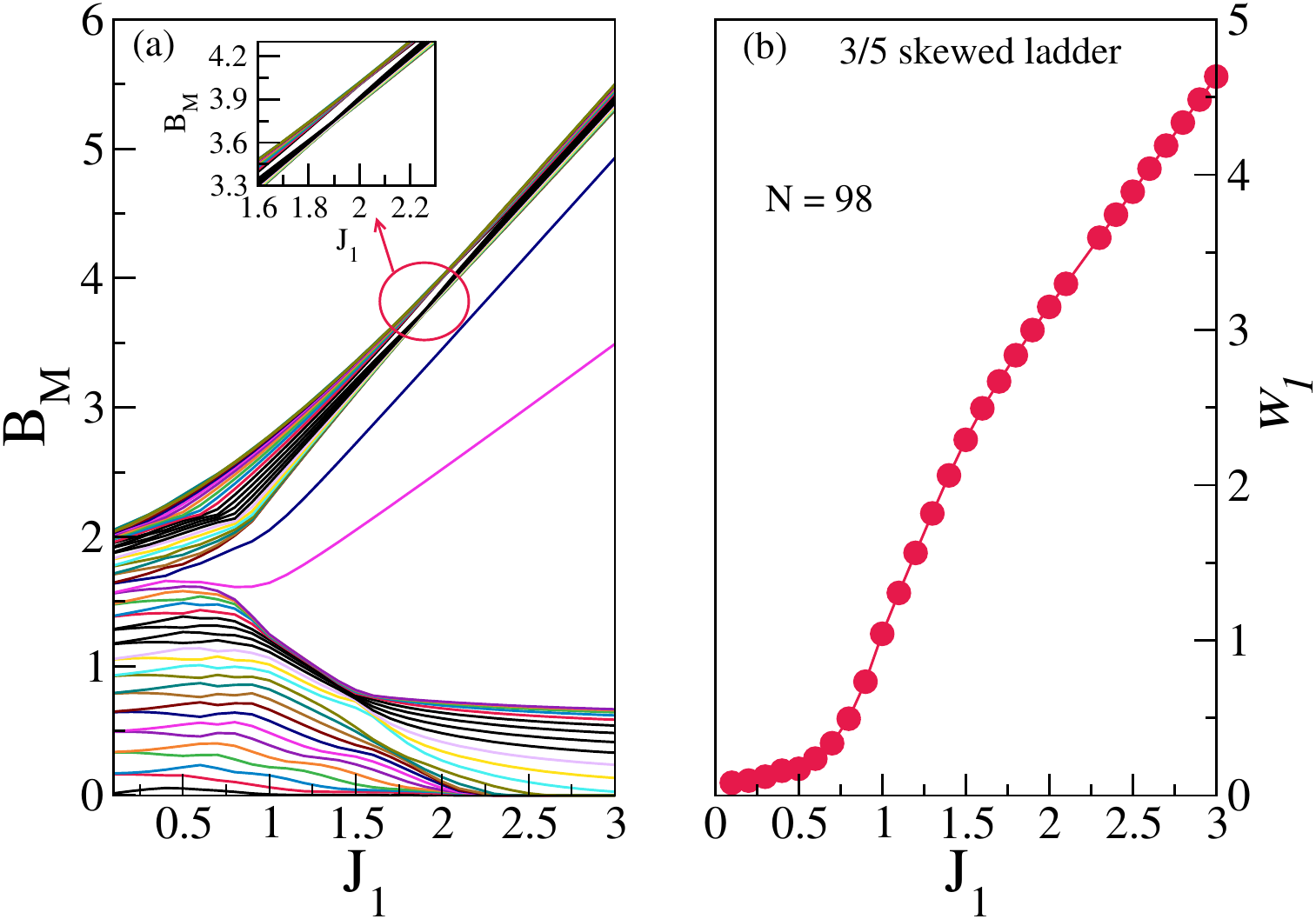}
	\caption{\label{fig:BM35}(a) The magnetic field $(B_M)$ required to close
	the energy gap between successive lowest energy $M_s$ states vs the rung
	bond interaction $(J_1)$ for a 3/5 skewed ladder. 
	The inset highlights the region $1.6 < J_1 < 2.3$, providing a closer view of a
        gap opens between the $B_M$ bands near the 3/4 plateau.
	(b) The width of the $m = 1/2$ plateau vs. $J_1$.}
	\end{center}
\end{figure}

\begin{figure}
\begin{center}
	\includegraphics[width=3.2 in]{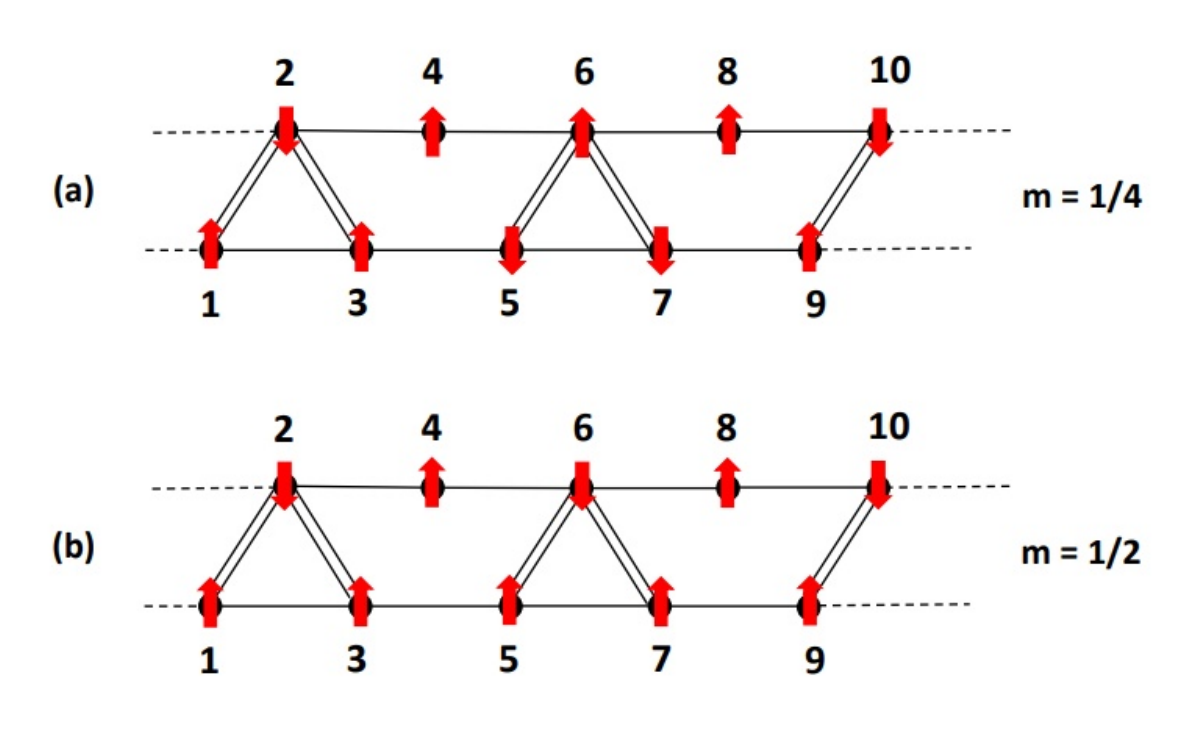}
	\caption{\label{fig:spr35}The arrangement of spins in a unit cell of a 3/5
	ladder at large $J_1$ limit is shown for (a) $m = 1/4$ gs and
	(b) $m = 1/2$ plateau state.}
	\end{center}
\end{figure}

\begin{figure}
\begin{center}
	\includegraphics[width=3.2 in]{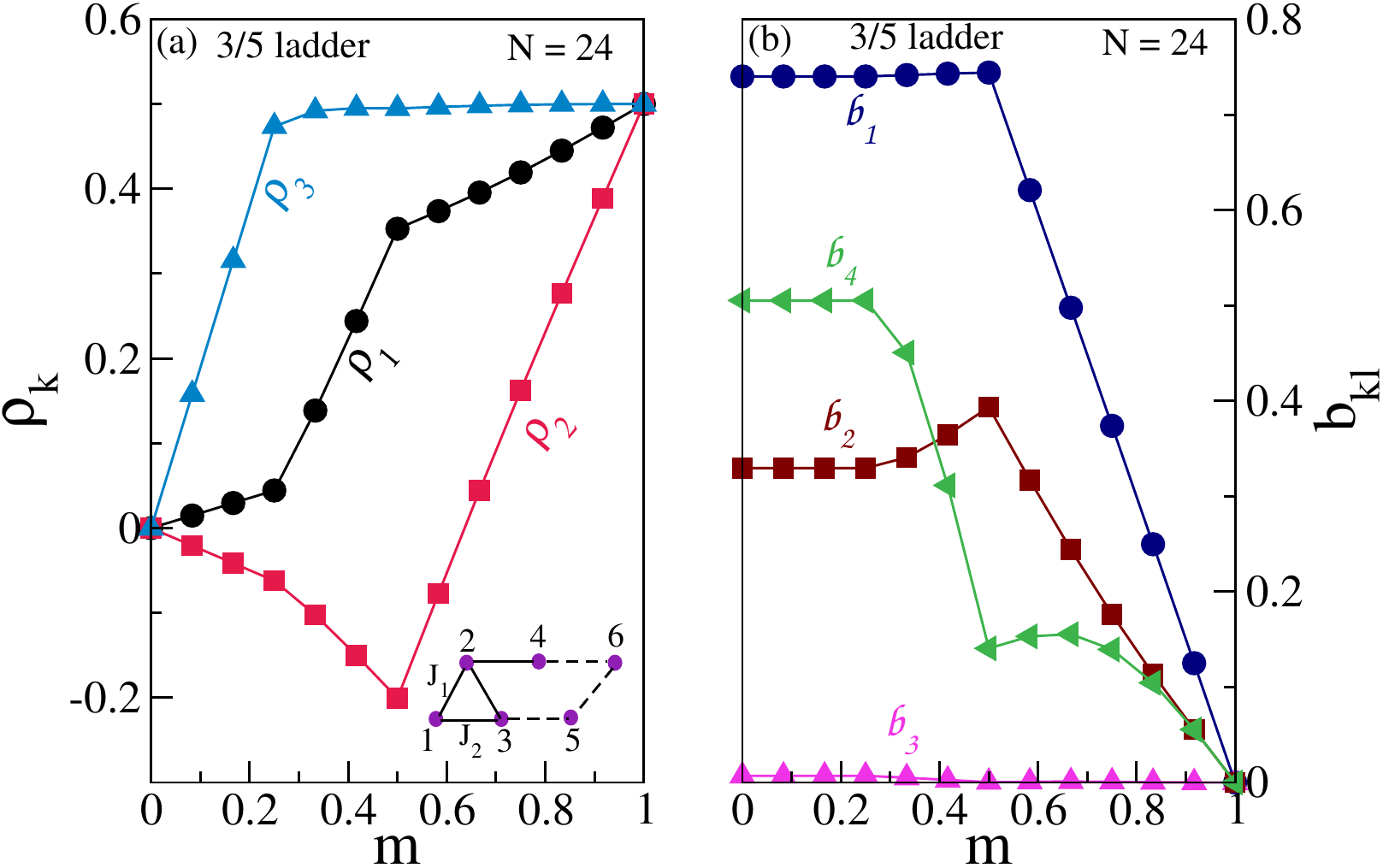}
	\caption{\label{fig:sdbo35}The spin densities and bond orders in a unit cell
	of a 3/5 ladder of $N =24$ sites with a periodic boundary condition.
	Here $\rho_1$ is the spin density at base sites of the triangle (1, 3), $\rho_2$ 
	is at apex site (2) and $\rho_3$ is at site 4. \urwchancery{b}$_1$ corresponds 
	to the rung bonds ($b_{12}$ and $b_{23}$), \urwchancery{b}$_2$ connects apex 
	of triangle and  the $4^{th}$ spin ($b_{24}$), \urwchancery{b}$_3$ is the bond 
	between sites at the base of the triangle ($b_{13}$) and \urwchancery{b}$_4$ 
	connects the sites 3 and 5 ($b_{35}$).}
	\end{center}
\end{figure}

In Fig.~\ref{fig:BM35}, $B_M$ is plotted as a function of $J_1$, and for $J_1 > 0.51$
$B_M$ forms two bands, lower band corresponds to 1/2 plateau, and upper
band corresponds to saturation magnetization. For $1.6 < J_1 < 2.3$ a gap opens
between $B_M$ bands near the 3/4 plateau. Fig.~\ref{fig:BM35}(b) gives the width of $m=1/2$ 
plateau, $w_1$, as a function of $J_1$ for $N=98$. $w_1$ is finite irrespective of the system 
size and $J_1$ and it increases slowly with $J_1$ up to $J_1 < 0.5$ and for $J_{1}>0.5$, $w_1$ 
shows a sharp and linear variation with large slope. The finite size effect of $w_1$ is almost 
negligible for large $J_1$. In each unit cell of the 3/5 ladder, three spins are on the triangle 
and one is attached to apex of the triangle. Therefore, there are only three unique sites: sites 1 
and 3 are equivalent and have spin density $\rho_1$, site 2 has spin density $\rho_2$ and site 4 
has spin density $\rho_3$ (Fig.~\ref{fig:schematic}(d)). There are four types of unique bonds: 
first type of bond \urwchancery{b}$_1$  are the rung bonds $b_{12}$, $b_{23}$, second type 
\urwchancery{b}$_2$ connects apex of triangle and  the $4^{th}$ spin, e.g. $b_{24}$ bond, the third type of bond 
\urwchancery{b}$_3$ is between 
sites at the base of the triangle e.g. $b_{13}$ and the fourth type of bond \urwchancery{b}$_4$ connects the 
base of two neighboring triangles e.g. $b_{35}$. The spin densities and bond orders are calculated 
as function of magnetization to understand the spin configuration of the gs as well as the plateau
phases for large $J_1$ (Fig.~\ref{fig:sdbo35}). We show the spin arrangements in the large $J_1$ 
limit, in  Fig.~\ref{fig:spr35} for the gs at $m = 1/4$ and $m=1/2$ plateau states.

For the $m=1/4$ plateau the rung bonds $b_{12}$  and $b_{23}$ form strong singlet, whereas, spins on
sites 1 and 3 interact ferromagnetically as shown pictorially in Fig.~\ref{fig:spr35}(a).
The $b_{24}$ bond has a weak anti-ferromagnetic alignment of spins. The spin densities are 
$\rho_{1}=0.044$, $\rho_{2}=-0.062$ and $\rho_{3}=0.5$. Spin densities and bond orders for 
$m = 1/2$ plateau state are shown in Fig.~\ref{fig:sdbo35}; $b_{12}$  and $b_{23}$ are strong 
singlet dimers, $b_{24}$ remains weakly antiferromagnetic while $b_{13}$ bond becomes 
ferromagnetic. Spin densities  at sites 1 and 3 have 0.353, whereas, these are $-0.20$ and 0.5 
at sites 2 and 4 respectively. Thus effectively one free spin-$\frac{1}{2}$ is contributed by the 
triangle and the other free spin-$\frac{1}{2}$ comes from site 4.
All spin densities increase linearly with $m$ for $m > 0.5$. For $J_1=5.0$,
\urwchancery{b}$_3$ is vanishingly small (0.005) for $m$ ranging between 1/4 and 1/2. 
\urwchancery{b}$_2$ increase from  0.33 to 0.394 as $m$ goes from 0 to 1/2 and it 
decreases thereafter. The first type bond \urwchancery{b}$_1$ nearly 1 implying a 
very strong singlet bond for $m$ up to 1/2, this bond becomes weak after $m=1/2$. Fourth type of bond 
\urwchancery{b}$_4$ is a weak ferromagnetic bond.

\section{\label{sec:QP} QP phase in 3/4 and 5/5 ladder}
In the previous section, we noted that both the 3/4 and 5/5 ladders exhibit elementary magnetic
steps of $\Delta M =2$ in the $m-B$ curve, while this feature is absent in the case of 3/5 ladder.
The steps of $\Delta M=2$ in the $m-B$ curve indicates $\Delta S =2$, the bi-magnon excitation in the system which 
characterises the quadrupolar (QP) or $n$-type of spin-nematic phase which does not break 
time-reversal symmetry~\cite{hikihara2008,kecke2007,meisner2006}. The general order parameter for this phase 
can be defined in terms of rank-2 tensor operator and written as~\cite{Penc2011},  
\begin{eqnarray}
	\hat{Q}_{kl}^{\alpha\beta} = \hat{S}_k^{\alpha}\hat{S}_l^{\beta} + \hat{S}_l^{\alpha}\hat{S}_k^{\beta} - \frac{2}{3}(\hat{S}_k \cdot \hat{S}_l)\delta_{\alpha\beta}
\label{Eq:Qab}
\end{eqnarray}
where $\alpha$ and $\beta$ represent the cartesian coordinates such as $x$, $y$ and $z$ and $k$ and $l$ are site indices. 
In these systems only the expectation value of the $x^2-y^2$ component of $\hat{Q}$  is finite and can be 
written as 
\begin{eqnarray}
	\hat{Q}^{x^2-y^2}_{kl}=\frac{1}{2}(\hat{S}^+_{k,l} \hat{S}^+_{k+1,l} + \hat{S}^-_{k,l} \hat{S}^-_{k+1,l}).
\label{Eq:Qx2y2}
\end{eqnarray}
The $\hat{Q}^{(x^2-y^2)}$ component shows quasi long range order, 
while other components are vanish ~\cite{hikihara2008,aslam_magnon}. In the thermodynamic limit,
this order parameter also goes to zero. Hence we call it quasi long range order. Another characteristic 
of this phase is that the quadrupolar order correlation decays slower than 
the spin-spin correlation. In the 3/4 skewed ladder, the effective spin-$\frac{1}{2}$ of each ring interacts 
ferromagnetically with the effective spin on the neighboring ring, resulting in a high-spin ground state. In contrast, the 5/5 ladder 
 exhibits antiferromagnetic interactions between the effective spins of adjacent rings, leading to a 
nonmagnetic ground state across the entire parameter range of $J_1$~\cite{geet}. 
The longitudinal spin-spin correlation function can be approximated by 
$A_{z}cos(2\pi \rho r)$, where $A_{z}$ is a constant~~\cite{hikihara2008}. We find the spin density has similar 
periodicity as the correlation function and also does not decay over short distances. Therefore 
we can fit the spin density dependence on distance to the function $A_{z}cos(2\pi \rho r)$. 
The cosine function represents the spiral nature of the spin density wave and $\theta$ is pitch 
angle between nearest spins. $\theta$ can be extracted from spin density calculation with OBC. 
Pitch angle can be easily calculated using the spin density which 
shows wave like behaviour for a given $M_s$ as shown in Fig.~\ref{fig:pitch34}(b) and 
Fig.~\ref{fig:pitch55}(b). If the wavelength of spin density wave is L then the pitch angle is given 
by $\frac{2 \pi}{L}$ and can be fitted to the expression~\cite{hikihara2008,aslam_magnon},
\begin{eqnarray}
	\frac{\theta}{\pi} = \rho =\frac{1}{q}(1-\frac{M}{M_\mathrm{max}}),
	\label{eq:rho}
\end{eqnarray}
where $M_\mathrm{max}$ is the saturation magnetization and $q = 2$ implies a quadrupolar phase.

Another important quantity of this phase is finite binding energy of two magnons 
condensate and the binding energy can be defined as ~\cite{vekua2007,aslam_magnon}
\begin{eqnarray}
	E_b = \frac{E_{0}(M+2)+E_{0}(M)-2E_{0}(M+1)}{2}.
	\label{eq:BE}
\end{eqnarray}
$E_{0}(M)$ is the lowest energy in the sector M; a finite negative
value of $E_b$ indicates that the simultaneous flipping of two spins to get the lowest 
energy state with $M_s=M+2$ from $M_s=M$ 
is energetically favorable compared to successively flipping one spin at a time. This manifests 
as steps of $\Delta M = 2$ in the $M-B$ curve. The attractive nature of two magnons leads to 
the formation of a two magnon bound state resulting in a quadrupolar phase. In this paper we characterize the 
quadrupolar phase using the finite binding energy between two magnons, steps of $\Delta M=2$ 
in magnetization and the linear variation of pitch angle with $m$.

\subsection{\label{sbsec:QP34} QP phase in the 3/4 ladder}
\begin{figure}
\begin{center}
        \includegraphics[width=3.0in]{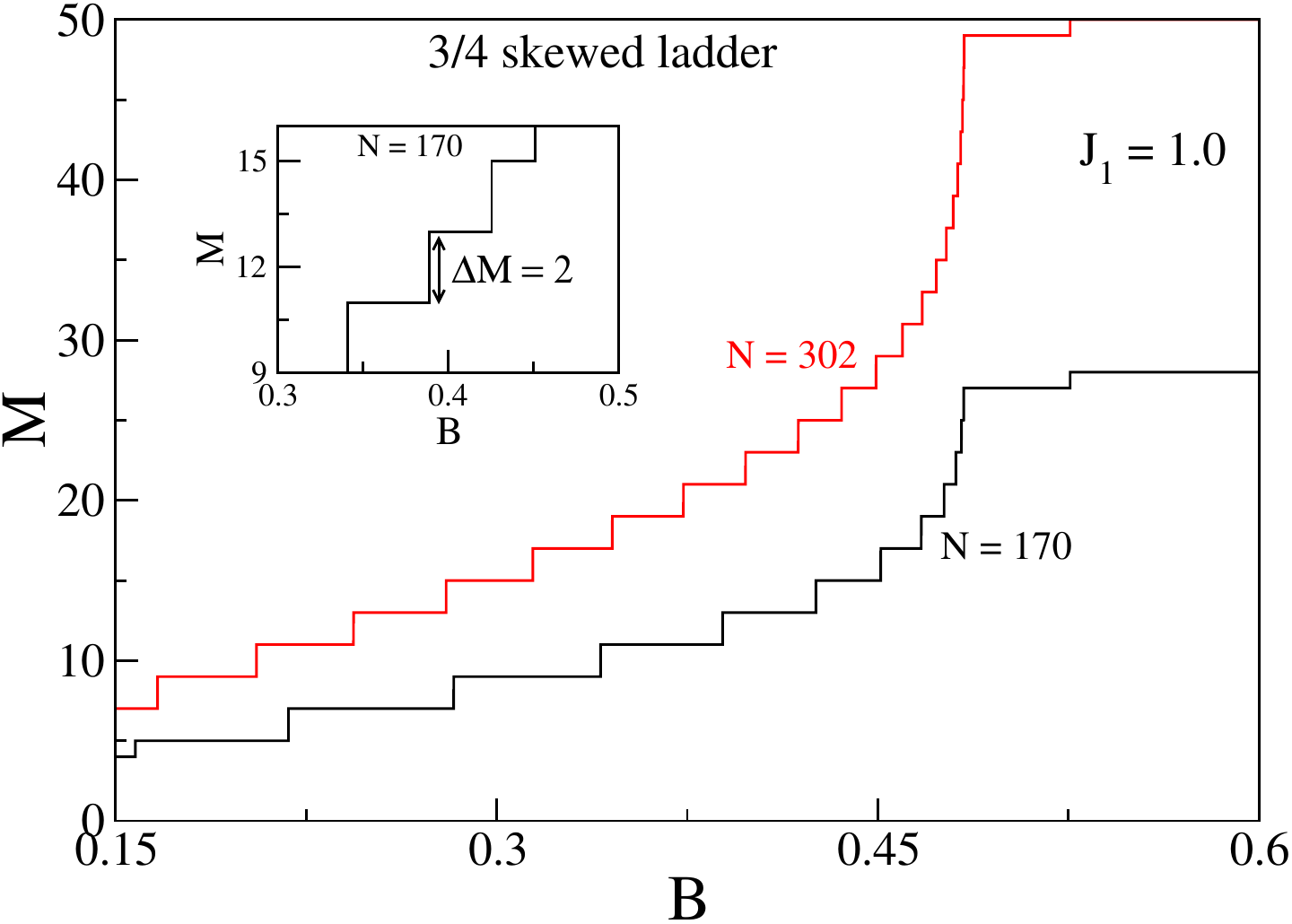}
	\caption{\label{fig:jump34} $M-B$ curves for a 3/4 ladder with $J_1 = 1.0$ showing 
        elementary magnetization steps of $\Delta M = 2$ for two system sizes N = 170 and 
	302 spins.The inset highlights the region
        $0.3 < B < 0.5$, providing a closer view of the
        magnetization steps of $\Delta M = 2$ in the $M-B$ curve for a system
        size N=170.}
        \end{center}
\end{figure}

\begin{figure}
\begin{center}
	\includegraphics[width=3.2in]{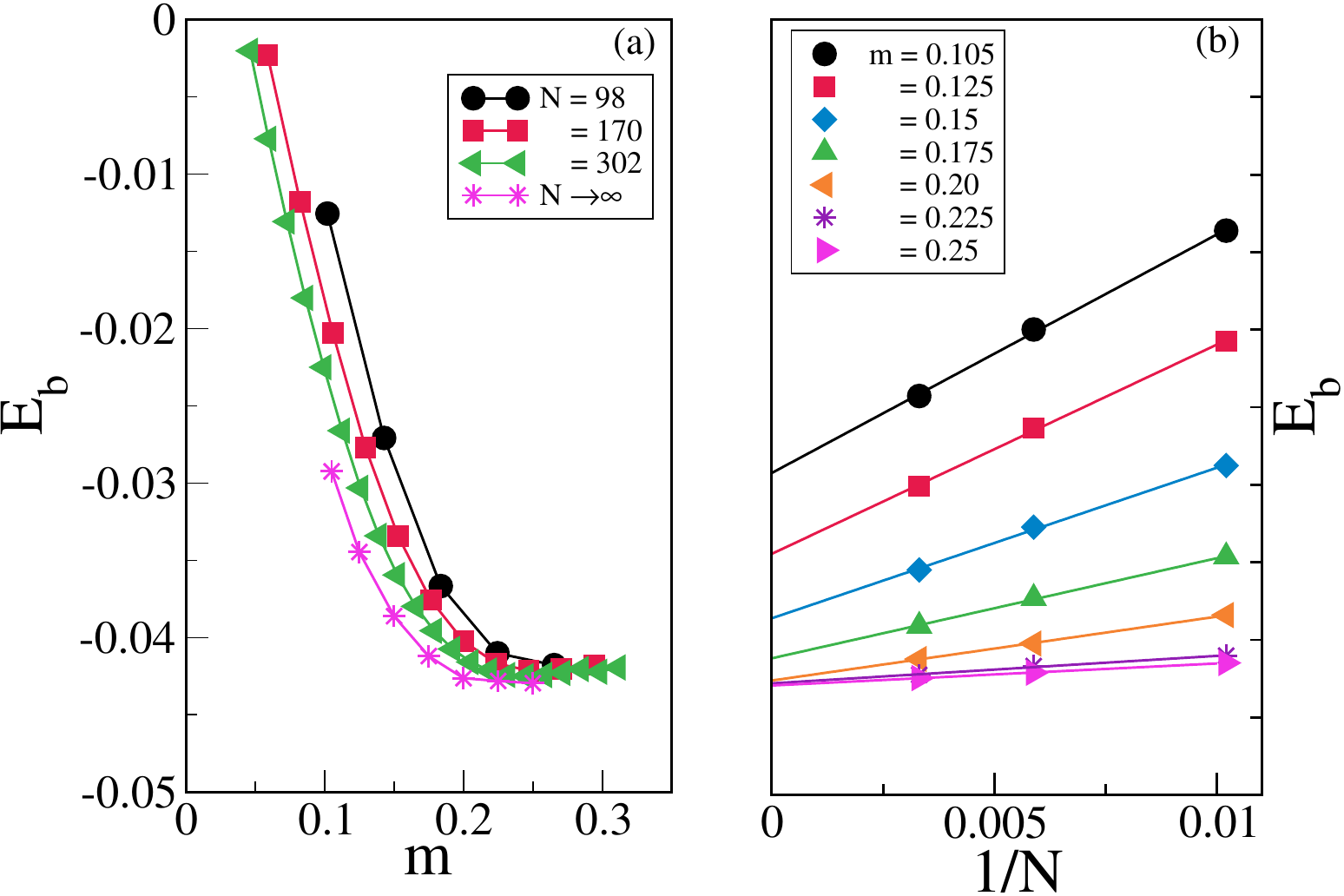}
	\caption{\label{fig:34BE}(a) The binding energy at different $m$ values for
	a 3/4 ladder before the $m=1/3$ plateau for different system sizes. The
	extrapolated binding energies are obtained from the linear fit of binding energy
	at different $m$ values with the inverse system size shown in (b). Scale on the vertical axis 
	is the same in both (a) and (b).}
	\end{center}
\end{figure}

The 3/4 ladder mimics the zigzag ladder with periodically missing bonds and one may expect the 
possibility of attractive interaction between the magnons due to the tendency of the system 
to transition into a ferrimagnetic state in certain parameter regime. In Fig.~\ref{fig:jump34}, 
the $M-B$ curves for two system sizes, $N=170$ and 302 spins show elementary 
magnetization steps of $\Delta M =2$ for $J_1=1.0$. 
The steps of $\Delta M =2$ remains restricted to magnetizations below 
the 1/3 plateau for $0.4 < J_1 < 1.58$ 
and it starts from $M$ between 5 and 7 in a system with OBC. The binding energy $E_b$ of the system 
as defined in Eq.~(\ref{eq:BE}) and it is plotted as a function of $m$ for $J_1=1.0$ in 
Fig.~\ref{fig:34BE}. We notice that the magnitude of $E_b$ increases with $m$ and it reaches a 
maximum around $m=0.2$ and decreases thereafter. For low values of $m$, $E_b$ has  dominant finite 
size effect and it extrapolates to a small value, whereas close to the 1/3 plateau the
finite size effect is small as shown in (Figs.~\ref{fig:34BE}(a) and~\ref{fig:34BE}(b)).

To understand the origin of the bound magnon pair, we compute the local binding energy in the 
QP state for unique bonds `$j$'. Unique bonds in the 3/4 ladder are the $1-2$ bond (`$j$'=1), 
the $1-3$ bond (`$j$'=2) and the $2-4$ bond (`$j$'=3). We define the bond energy 
${\Delta}_{j}^{T/L}$, where T(L) are the transverse (longitudinal) bond operators 
${\hat{b}_{j}}^{T/L}$ = ${\hat{S}_{k_j}}^{T/L}$$\cdot$${\hat{S}_{l_j}}^{T/L}$ where $\hat{S}^{T}$ and $\hat{S}^{L}$ 
represent the longitudinal and transverse component of the spin 
operators, and $k$ and $l$ are the site indices of the bond $j$. The local binding energy 
$\Delta_{j}^{T/L}$ of the $j^{th}$ bond is given by 
\begin{eqnarray}
        \Delta_{j}^{T/L}(M) = \frac {1}{2} \left[\langle{b_{j}^{T/L}(M+2)}\rangle+\langle{b_{j}^{T/L}(M)}\rangle \right. \nonumber \\
\qquad \qquad \qquad \qquad \qquad \quad \left. -2\langle{b_{j}^{T/L}(M+1)}\rangle\right]
\label {eq:bindingenergy}
\end{eqnarray}
where the expectation values are for the lowest energy state in the specified magnetization sector.

In table~\ref{tab:eb34}, $\Delta_{j}^{T/L}$ are presented for 3/4 ladder with $N=24$ sites in 
$M_s=2$ sector. The longitudinal component of the leg bonds ($2-4$) connecting two neighboring 
triangles have highest contribution to the two magnon binding energy while the longitudinal 
component of the rung bond ($1-2$) gives the second highest contributor as shown in 
table~\ref{tab:eb34}. The least contribution comes from the transverse component of the bond 
forming the base of the triangles ($1-3$). We note that the overall contribution from the 
longitudinal components is negative while the overall contribution from the transverse components 
is positive. After taking into account both the longitudinal and transverse components we observe 
that the contribution from all the three bond types are negative. The major contribution of binding 
energy comes due to effective ferromagnetic exchange between effective spin between two consecutive 
triangles.    
\begin{table*}
    \centering
    \begin{tabular}{c @{\hspace{1.4cm}} c @{\hspace{1.4cm}} c @{\hspace{1.4cm}} c @{\hspace{1.4cm}} c @{\hspace{1.4cm}} c}
        \hline
        $J_1$  & Bond Index (j)  & $n_j$  & $\Delta_j^L (M_s=2)$  & $\Delta_j^T (M_s=2)$  & $n_j \times (\Delta_j^L + \Delta_j^T)$ \\
        \hline
         & 1 & 4 & $-0.00362675$ & \, \,$0.00222080$ & $-0.00562380$ \\
        1.0 & 2 & 2 & \, \, $0.00118497$ & $-0.00310089$ & $-0.00383184$ \\
         & 3 & 4 & $-0.00803502$ & \, \,$0.00439510$ & $-0.01455970$ \\
        \hline
        & & \multicolumn{4}{c}{\hspace{3.4cm} Binding energy per unit cell = $-0.02401534$} \\
	\hline
    \end{tabular}
    \caption{\label{tab:eb34}The binding energy for the unique bonds in a unit cell of a 3/4
    skewed ladder of $N = 24$ spins with PBC at $J_1 = 1.0$. Here $n_j$ is the number of unique bonds
    per unit cell. The contribution of the transverse ($\Delta_j^T$) and longitudinal ($\Delta_j^L$)
    binding energies are shown separately. The numbers in the $\Delta_j^L$ and $\Delta_j^T$ columns
    show the contribution to binding energy per single bond. The last column shows the contribution
    from different unique bond types in a unit cell.}
\end{table*}
\begin{figure}
\begin{center}
	\includegraphics[width=3.2in]{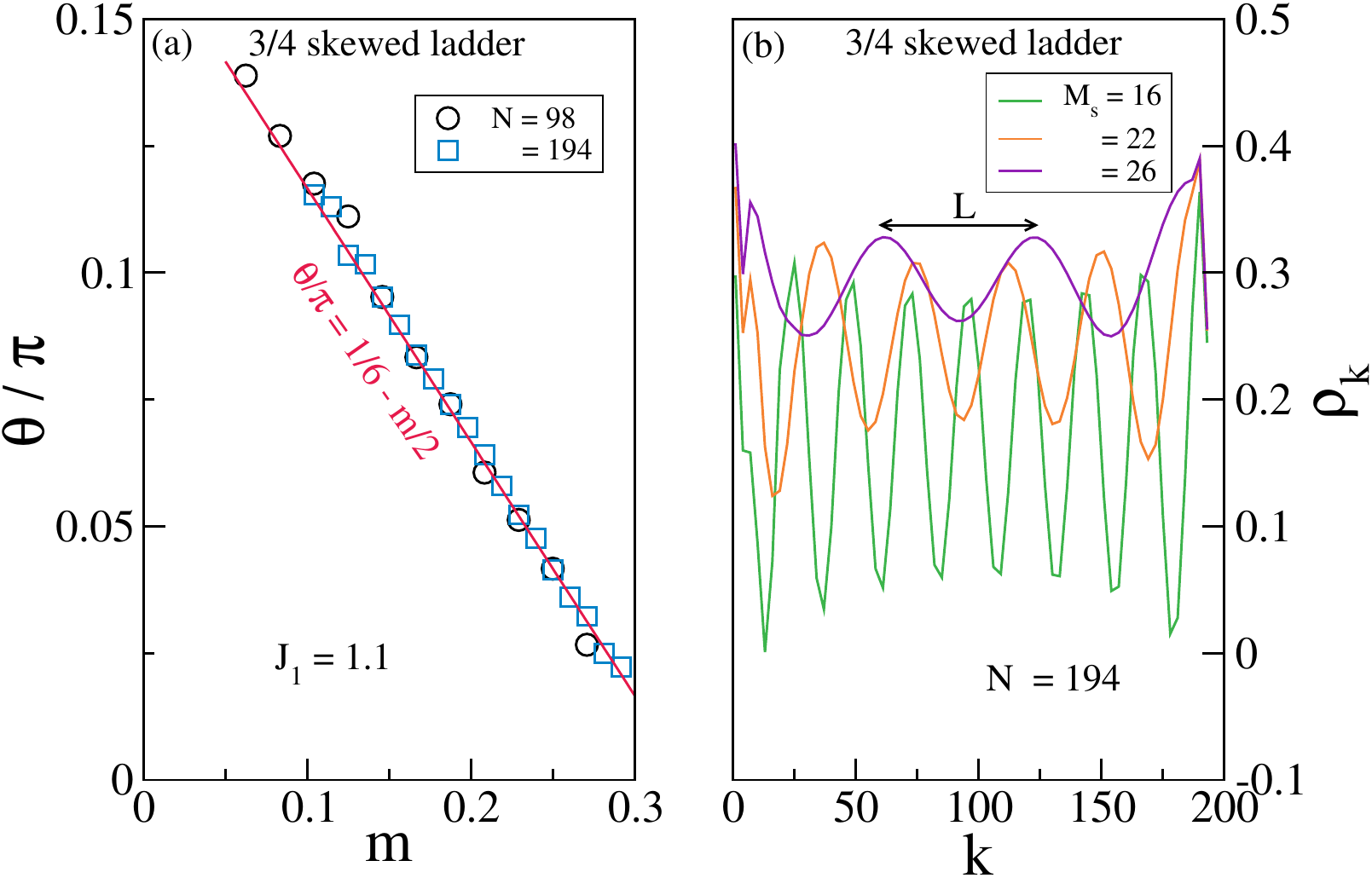}
	\caption{\label{fig:pitch34}(a) The linear behavior of the pitch angle with
	the magnetization of the 3/4 ladder before the 1/3rd plateau is shown
	for $N = 98$ and 194. (b) The variation of spin density for three different
        $M_s$ sectors are shown for a system of $N = 194$ spins. L is 
	the wavelength of the spin density wave.}
	\end{center}
\end{figure}
The third evidence of the quadrupolar phase is the linear variation of pitch angle
$\theta$ with $m$ (Eq.~\ref{eq:rho}). In Fig.~\ref{fig:pitch34}(a), the $\theta/\pi$ is plotted as a function of $m$
for two system sizes with 98 and 194 spins (circles and squares respectively), for $J_1 = 1.1$. The variation 
of the spin densities in systems with OBC are shown in Fig.~\ref{fig:pitch34}(b). $\theta$ is calculated
from the spin density wave using the relation $\theta = \frac{2\pi}{L}$, 
and these values can be
fitted with the relation $\frac{\theta}{\pi}=\frac{1}{q}(\frac{1}{3}-m)$ with $q=2$. Similar pattern is observed for various values
of  $0.5 < J_1 < 1.5$.

\subsection{\label{sbsec:QP55}QP phase in the 5/5 ladder}
\begin{figure}
\begin{center}
        \includegraphics[width=2.7in]{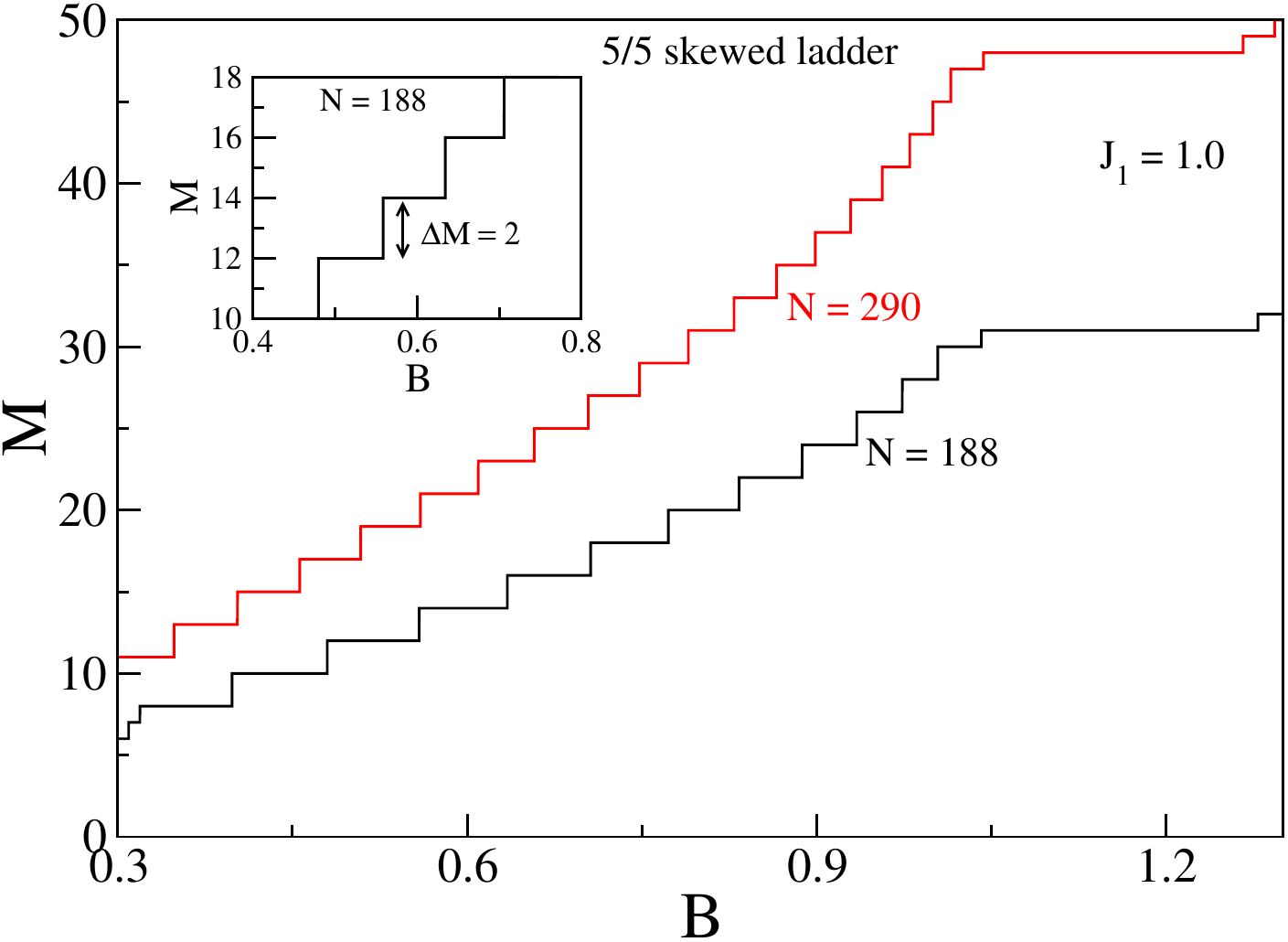}
        \caption{\label{fig:jump55} $M-B$ curves for a 5/5 ladder with $J_1 = 1.0$ showing
        elementary magnetization steps of $\Delta M = 2$
        in magnetization for two system sizes $N = 188$ and 290 spins. The inset highlights
        the region $0.4 < B < 0.8$, providing a closer view of the
        magnetization steps of $\Delta M = 2$ in the $M-B$ curve for a system size $N = 188$.}
        \end{center}
\end{figure}
\begin{figure}
\begin{center}
        \includegraphics[width=3.2in]{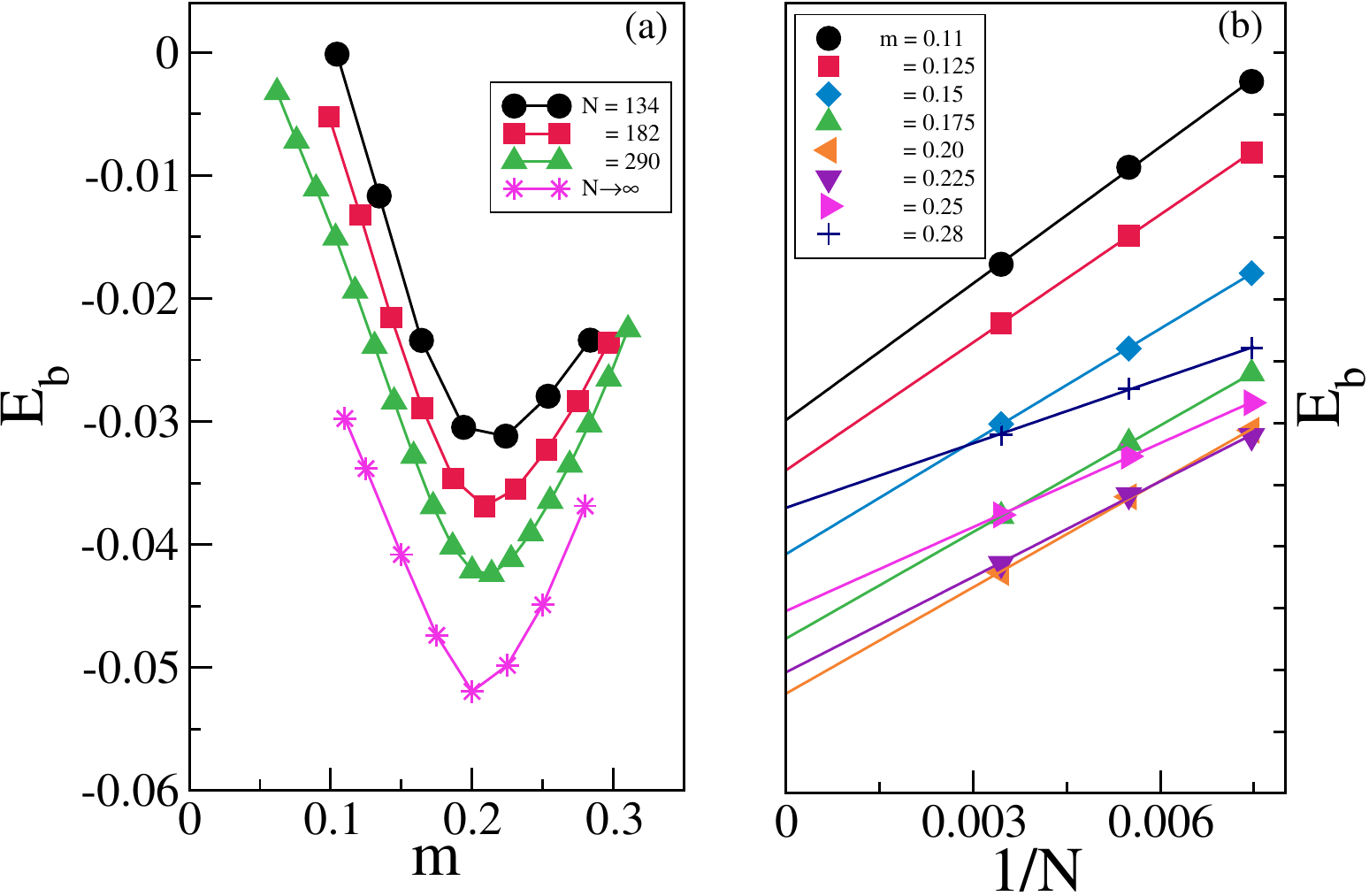}
        \caption{\label{fig:55BE}(a) The binding energy at different $m$ value for
	the 5/5 ladder below the $m=1/3$ plateau for different system sizes. (b) The
        extrapolated binding energies are obtained from a linear fit of binding energy
	for each $m$ value with the inverse system size. Scale on the vertical axis is the same
	in both (a) and (b).}
        \end{center}
\end{figure}

A similar analysis is carried out for 5/5 ladder shown in Fig.~\ref{fig:schematic}(c) and in
Fig.~\ref{fig:jump55}, $M-B$ curves for two system sizes $N=188$ and 290 show the magnetization steps
of $\Delta M = 2$ at $J_1 = 1.0$. Similar to the 3/4 ladder, the
magnetization steps of $\Delta M = 2$ in $M-B$ curve remain
restricted to $m$ values below the $m = 1/3$ plateau for the parameter range  $0.4 < J_1 < 1.2$
and it start from $M$= 7 to 9 in a system with OBC and depends weakly on system size. The binding
energy $E_b$ is plotted as a function of $m$ for $J_1=1.0$ in Fig.~\ref{fig:55BE}(a). We notice
that the magnitude of $E_b$ increases with $m$ and it reaches a maximum around $m=0.2$ and
decreases afterwards. In this system, similar to the 3/4 ladder, for small values of $m$, $E_{b}$
shows dominant finite size effect and extrapolates to small values, whereas close to the
1/3 plateau, the finite size effect is small  (Fig.~\ref{fig:55BE}(b)).
%
In the 5/5 ladder, there are only two unique sites and three unique bonds
and in table~\ref{tab:eb55}, the $\Delta_{j}^{T/L}$ are presented for $N=24$ sites in $M_s=2$ sector
with PBC. Contribution of various per bond binding energies contributing to the total
$\sum_{n_j} \left( \Delta_j^{T} + \Delta_j^{L} \right)$ in the 5/5 ladder is shown
in table~\ref{tab:eb55}, here $n_j$ represents the number of `$j$' type of bonds
in a unit cell; `$j$'=1 is the $1-2$ bond, `$j$'=2 is the $1-3$ bond
and `$j$'=3 is the $2-4$ bond. The $\Delta_{1}^T$ and $\Delta_{1}^L$ both are negative
and the transverse component has the highest absolute value,
the longitudinal component of the $2-4$ type bonds give the second highest contribution. Both
the longitudinal and the transverse components of $1-3$ type bonds  give positive contribution
to the total binding energy. In a unit cell, there are four $1-3$ type bonds,
resulting in a high positive contribution from this kind. The substantial positive contribution
from the $j=2$ $(1-3)$ bond type cancels out the overall negative contribution from the $j=1$ $(1-2)$
and $j=3$ $(2-4)$ bond types, resulting in a low binding energy per unit cell.
Similar to 3/4 skewed ladder, overall contribution from the longitudinal components is negative while the
overall contribution from the transverse components is positive.
\begin{figure}
\begin{center}
        \includegraphics[width=3.2in]{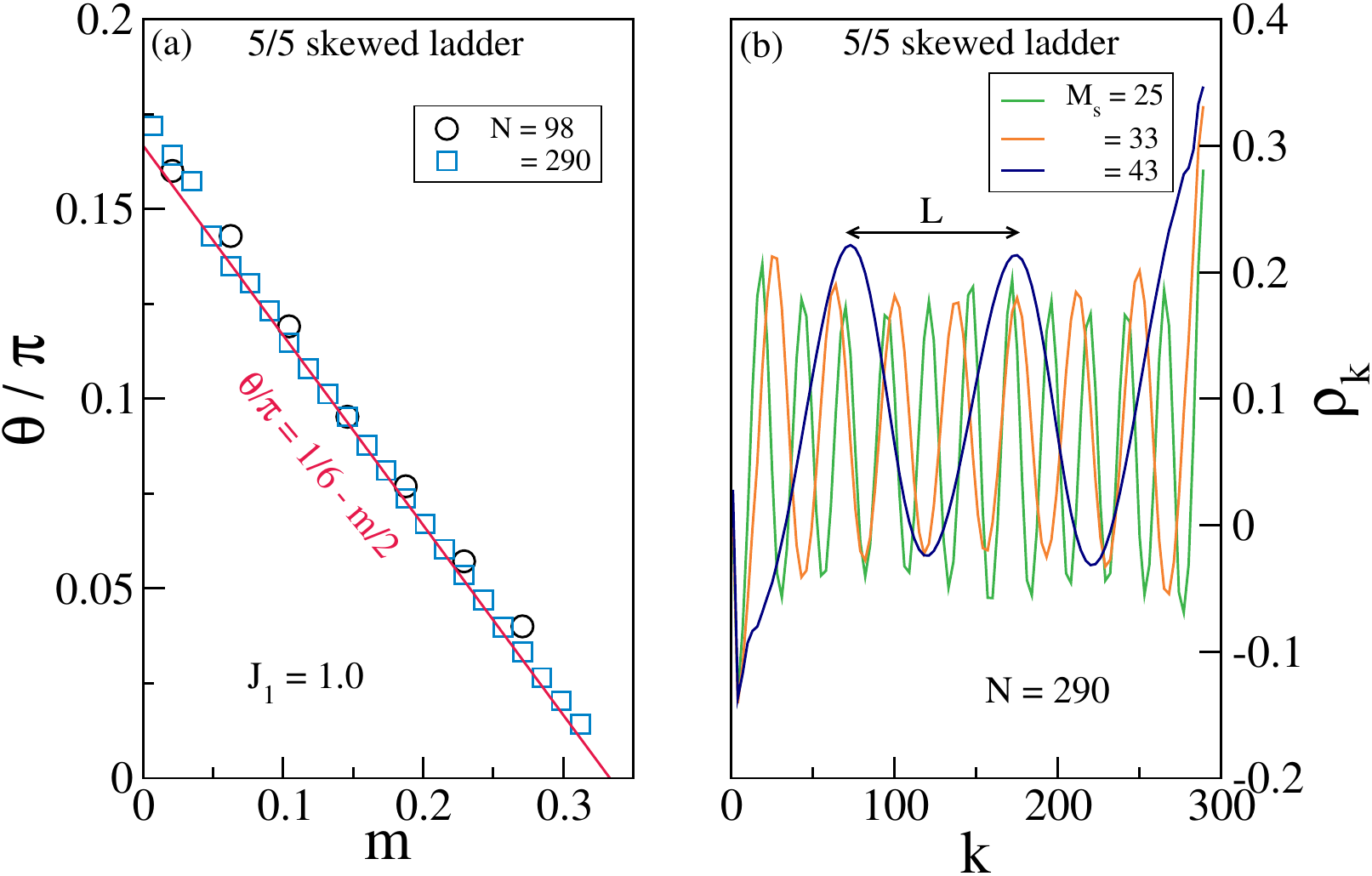}
	\caption{\label{fig:pitch55}(a) The linear behavior of the pitch angle with
        magnetization of the 5/5 ladder before the 1/3rd plateau is shown
	for $N = 98$ and $N = 290$ spins. (b) Variation of the spin densities
	is shown for three $M_s$ sectors of $N = 290$ spins system. L is the
	wavelength of the spin density wave.}
	\end{center}
\end{figure}
The linear variation of pitch angle $\theta$ is shown in Fig.~\ref{fig:pitch55}(a)
and the $\theta/\pi$ is plotted as a function of $m$ for the two system sizes, $N=98$
and 290 spins, respectively, for $J_1 = 1.0$. The circles and squares represent
system sizes $N=98$ and 290, respectively. The variation of
spin densities in a system of N = 290 with OBC are shown in Fig.~\ref{fig:pitch55}(b). $\theta$ is calculated from the
spin density wave using the relation $\theta = \frac{2\pi}{L}$,
and these values can also be fitted to the straight line $\frac{\theta}{\pi}=\frac{1}{q}(\frac{1}{3}-m)$ where $q=2$.
\begin{table*}[t]
    \centering
    \begin{tabular}{c @{\hspace{1.4cm}} c @{\hspace{1.4cm}} c @{\hspace{1.4cm}} c @{\hspace{1.4cm}} c @{\hspace{1.4cm}} c}
        \hline
        $J_1$   & Bond Index (j) & $n_j$ & $\Delta_j^L (M_s=2)$ & $\Delta_j^T (M_s=2)$ & $n_j \times (\Delta_j^L + \Delta_j^T)$ \\
        \hline
         & 1 & 2 & $-0.01462905$ & $-0.02794142$ & $-0.0851409$ \\
        1.0 & 2 & 4 & \, \, $0.01398213$ & \, \,$0.02948677$ & \, \,$0.1738760$ \\
         & 3 & 2 & $-0.02578724$ & $-0.02226299$ & $-0.0961005$ \\
        \hline
        & & \multicolumn{4}{c}{\hspace{3.4cm} Binding energy per unit cell = $-0.0073654$} \\
        \hline
    \end{tabular}
    \caption{\label{tab:eb55}The binding energy for the unique bonds in a unit cell of a 5/5
        skewed ladder of $N = 24$ spins with PBC at $J_1 = 1.0$. Here $n_j$ is the number of unique
        bonds per unit cell. The contribution of the transverse ($\Delta_j^T$) and longitudinal
        ($\Delta_j^L$) binding energies are shown separately. The numbers in the $\Delta_j^L$ and
        $\Delta_j^T$ columns show the contribution to binding energy per single bond.
        The last column shows the contribution from different unique bond types in a unit cell.}
\end{table*}
\section{\label{sec:disc} Summary and conclusions}
In this paper, a spin-$\frac{1}{2}$ isotropic Heisenberg model on three types of skewed ladders,
namely, 3/4, 5/5 and  3/5 is studied in the presence of Zeeman
magnetic field $B$.
These systems show interesting magnetization plateaus, besides the 3/4 and 5/5 ladders show emergent quadrupolar
phase. We have numerically solved these models in Eqs.~(\ref{eq:ham34}),~(\ref{eq:ham55})
and~(\ref{eq:ham35}) by employing the ED and the DMRG numerical methods. We calculate the plateau
width and predict the dominant spin configuration in the plateau states based on spin density
and bond order calculations. The QP phase is characterized by using the
steps of $\Delta M =2$ in the $M-B$ curve, finite binding energies and linear variation of the pitch angle $\theta$ with
$m$. To the best of our knowledge the ladders 3/4 and 5/5 are unique systems in which both
plateau and QP phases can be stabilized.

In the spin-$\frac{1}{2}$ system on the 3/4 skewed ladder there are six spins per unit cell,
OYA condition~\cite{OYA_97} suggests the possible plateau states at $m = 0$, 1/3, 2/3 and 1
whereas, our system shows plateaus only at 1/3, 2/3 and 1. The plateau at 1/3 of this system
is similar to that seen in a zigzag ladder~\cite{meisner2007, okunishi_prb2003, okunishi_jpsj2003},
but the plateau at 2/3 is unique to the ladder system. For $J_1 > 1.58$ the 1/3 plateau
becomes the gs even in the absence of $B$. For the 5/5 skewed ladder with six spin-$\frac{1}{2}$
objects per unit cell, OYA rule again predicts plateaus at $m = 0$, 1/3, 2/3 and 1 magnetization.
Even though the OYA condition is only a necessary condition, we find calculated values of the plateaus
are indeed consistent with the values predicted by the OYA condition. We also note that only
the 1/3 plateau is dominant with large width. Other plateaus are weak and have vanishingly small widths.
In the third system considered here, the 3/5 skewed ladder there are four sites per unit cell and  the
enlarged magnetic unit cell predicts plateaus at $m = 0$, 1/4, 1/2, 3/4 and 1. we observe the plateaus only
at $m$ = 1/4, 1/2, 3/4 and 1. However, only 1/2 plateau has large width; other plateaus are restricted
to small parameter regime and have very small widths. In the large $J_1 (> 2.3)$ limit, the gs is a ferrimagnetic with $m=1/4$.

The HAF spin-$\frac{1}{2}$ model on 3/4 and 5/5 ladder geometries exhibit QP phase besides
magnetization plateaus. Interestingly, this phase exists for low magnetic fields or $m$ below
1/3 which is very different from ferromagnetic $J_1-J_2$ spin-$\frac{1}{2}$ model where it exists
only at large magnetization or high magnetic field $B$ \cite{hikihara2008, lauchli2009}. In both
the systems $\theta$ vs $m$ plots show linear variation and have a slope of -1/2; irrespective of
the structural differences, the nature of $\theta-m$ behavior remains the same. The $E_b$ in
these systems is about half that found in the ferromagnetic spin-$\frac{1}{2}$ $J_1-J_2$
model~\cite{aslam_magnon}.

There are many open questions like; are these systems quantum spin liquids? If yes, what kind of
topological order do exist in these systems? What are the transport properties of these systems?
In summary, we have studied exotic phases in the 3/4, 5/5, and 3/5 skewed ladder systems
in the presence of a Zeeman magnetic field, and we observed that all three magnetic systems exhibit plateau phases.
In the 3/4 and 5/5 systems QP phase is stabilized at low magnetic field which is unique to these systems.

\ack
S.R. acknowledges the Indian National Science Academy and DST-SERB for supporting this work. M.K. acknowledges the SERB for financial 
support through Project File No. CRG/2020/000754.

\section*{Data Availability}
The data that support the findings of this study are available from the corresponding author upon reasonable request.

\section*{ORCID iDs}
Sambunath Das\includegraphics[width=8pt]{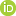}\href{https://orcid.org/0000-0003-3127-041X}{https://orcid.org/0000-0003-3127-041X} \\
Dayasindhu Dey\includegraphics[width=8pt]{./orcid.png}\href{https://orcid.org/0000-0002-5303-4146}{https://orcid.org/0000-0002-5303-4146} \\
Manoranjan Kumar\includegraphics[width=8pt]{./orcid.png}\href{https://orcid.org/0000-0002-7624-8155}{https://orcid.org/0000-0002-7624-8155} \\
S. Ramasesha\includegraphics[width=8pt]{./orcid.png}\href{https://orcid.org/0000-0001-8615-6433}{https://orcid.org/0000-0001-8615-6433}

\setcounter{figure}{0}
\setcounter{table}{0}
\renewcommand{\theequation}{A\arabic{equation}}
\renewcommand{\thefigure}{A\arabic{figure}}

\section*{References}
\bibliographystyle{iopart-num}
\bibliography{bibliography}
\end{document}